# Unexpected coexisting solid solutions in the quasi-binary Ag$^{(II)}$F$_2$/Cu$^{(II)}$F$_2$ phase diagram

D. Jezierski,[a] K. Koteras,[a] M. Domański,[a] P. Połczyński,[b] Z. Mazej,*[c] J. Lorenzana,*[d] W. Grochala*[a]

[a] Centre of New Technologies, University of Warsaw, 02097 Warsaw, Poland
[b] Faculty of Chemistry, Biological and Chemical Research Centre, University of Warsaw, 02089 Warsaw, Poland
[c] Department of Inorganic Chemistry and Technology, Jožef Stefan Institute, 1000 Ljubljana, Slovenia
[d] Institute for Complex Systems (ISC), Consiglio Nazionale delle Ricerche, Dipartimento di Fisica, Università di Roma "La Sapienza", 00185 Rome, Italy

* zoran.mazej@ijs.si , jose.lorenzana@cnr.it , w.grochala@cent.uw.edu.pl

## Abstract

High-temperature solid-state reaction between orthorhombic AgF$_2$ and monoclinic CuF$_2$ (y = 0.15, 0.3, 0.4, 0.5) in a fluorine atmosphere resulted in coexisting solid solutions of Cu-poor orthorhombic and Cu-rich monoclinic phases with stoichiometry Ag$_{1-x}$Cu$_x$F$_2$. Based on X-ray powder diffraction analyses, the mutual solubility in the orthorhombic phase (AgF$_2$:Cu) appears to be at an upper limit of Cu concentration of 30 mol % (Ag$_{0.7}$Cu$_{0.3}$F$_2$), while the monoclinic phase (CuF$_2$:Ag) can form a nearly stoichiometric Cu : Ag = 1 : 1 solid solution (Cu$_{0.56}$Ag$_{0.44}$F$_2$), preserving the CuF$_2$ crystal structure. Experimental data and DFT calculations showed that AgF$_2$:Cu and CuF$_2$:Ag solid solutions deviate from the classical Vegard's law. Magnetic measurements of Ag$_{1-x}$Cu$_x$F$_2$ showed that the Néel temperature (T$_N$) decreases with increasing Cu content in both phases. Likewise, theoretical DFT+U calculations for Ag$_{1-x}$Cu$_x$F$_2$ showed that the progressive substitution of Ag by Cu decreases the magnetic interaction strength |J$_{2D}$| in both structures. Electrical conductivity measurements of Ag$_{0.85}$Cu$_{0.15}$F$_2$ showed a *ca.* 2-fold increase in specific ionic conductivity (3.71·10$^{-13}$ ± 2.6·10$^{-15}$ S/cm) as compared to pure AgF$_2$ (1.85·10$^{-13}$ ± 1.2·10$^{-15}$ S/cm), indicating the formation of a vacancy- or F adatom-free metal difluoride sample.

## Introduction

Inspection of phase diagrams of binary and ternary compositions readily reveals that a huge share of chemical elements and simple chemical systems are immiscible with each other in the solid state in a nearly entire composition window. Since such systems are prone to phase separation their solid solutions do not form. Such behaviour strongly contrasts with that of the world of minerals, where extremely complex stoichiometries may be found. For example, the chemical composition of a common mineral pyrochlore, found in d'Oka (Canada), may be approximated as: (Na$_{2.35}$Ca$_{11.01}$Sr$_{0.14}$Mn$_{0.032}$Mg$_{0.11}$Ce$_{1.36}$La$_{0.27}$Nd$_{0.34}$Y$_{0.016}$U$_{0.064}$Th$_{0.13}$Te$_{0.016}$)(Nb$_{11.62}$Ta$_{0.3}$Ti$_{2.85}$Zr$_{0.24}$Fe$_{0.75}$)(O$_{49.25}$(OH)$_{0.41}$F$_{6.34}$), with certainly more constituting elements occurring at much smaller concentrations[1]. The fact that this Ca$^{2+}$$_2$Nb$^{5+}$$_2$O$_7$-type mineral adopts a high-symmetry "simple" *F*d-3m (No. 227) space group implies that many different cations (and even those of varying valences), occupy the same position in the crystal structure. In the case of the said pyrochlore, monovalent sodium coexists at one Wyckoff position with divalent alkali earth metals and transition metals, trivalent lanthanides, tri- and tetravalent actinides, and even post-transition semimetal from Group 16, tellurium. Many more similar cases are known. Pre-eminence of such substances, which formed from a melt at a certain moment in the history of Earth, suggests that miscibility may show a substantial synergy; for example, if a cation A is too different in size from a cation C, and they do not occur at the same crystallographic position in a quasi-binary A-ligand/C-ligand system, the addition of a cation B of intermediate size may facilitate the formation of a solid solution between all three components. This observation has laid ground for a renaissance of multicomponent solid solutions, labelled as "high entropy alloys" which have been first prepared in 2004[2]. Initially of importance mostly in materials engineering, such systems expanded in 2015 and later to more complex systems, notably oxides, further testifying to their enormous usefulness in many areas of technology[3]. After over two decades of systematic development, high entropy solid solutions now constitute a burgeoning field of science.

Metal fluorides constitute an important class of materials, which find use as catalysts, ionic conductors, active cathode materials for battery cells, in superconducting devices, as corrosion protection materials *etc.*[4–7]. However, the number of genuine fluoride solid solutions prepared to date is rather modest. During the last decade, several mixed-cation ionic Group 2 metal fluorides have been prepared, *i.e.* (Mg$_{1-x}$M$_x$)F$_2$ for M = Ca$^{2+}$, Sr$^{2+}$, Ba$^{2+}$ [8,9], and (M$^a$$_x$M$^b$$_{1-x}$)F$_2$ for M$^{a,b}$ = Ca$^{2+}$, Sr$^{2+}$, Ba$^{2+}$[10,11], largely with the intent to increase ionic mobility[12]. Only recently, multi-cation rutile-type fluorides, containing 4-7 cations (Co, Cu, Mg, Ni, Zn, Mn, and Fe) have been synthesized using a mechanochemical approach[13]. Metal fluorides are intriguing candidates for cathode material in the fluoride ion batteries[5,14]. Much effort has been made to improve their reversibility, with (Cu,M)F$_2$ (M=Fe, Ni) solid solutions showing a desirable performance[15,16]. However, to the best of our knowledge, only one example is known of a system with cations coming from two distinct periods of the Periodic Table. $Pd_{1-x}Zn_xF_2$ (with x varying in a broad range from 0.1 to 0.95), known since 1977, hosts paramagnetic high-spin $Pd^{2+}$ partly substituted by diamagnetic $M=Zn^{2+}$[17]. At the first sight, $(A,M)F_3$ (A=Ti, V; M=Zr, Hf)[18] constitute yet another examples of such solid solutions. However, only in the case of $Pd_{1-x}Zn_xF_2$ both cations show identical valence while $Pd_{1-x}Zn_xF_2$ is indeed isostructural with its $PdF_2$ predecessor. On the contrary, $(A,M)F_3$ (A=Ti, V; M=Zr, Hf) cannot be considered as solid solutions of the parent $AF_3$ and $MF_3$, but rather as salts of the type $A^{2+}M^{4+}F_6$, which do not resemble structurally neither $AF_2$ nor $MF_4$. Rarity of the true fluoride solid solutions hosting transition metal cations from two different periods of the Periodic Table, notably from the IV[th] one and the V[th] or the VI[th] ones, has to do with substantial differences in both ionic radii and chemistry of these lighter *vs.* heavier elements in general.

Thus, the compositional space of such solid solutions is a huge *terra incognita*. Driven partly by that, and partly by reasons outlined below, we describe our preliminary attempts to prepare solid solutions of $AgF_2$ and $CuF_2$, as a first step towards a hypothetical multicomponent high-entropy fluoride. Just like in the described cases of Ti and Hf, for example, the chemistry of Cu and Ag is vastly different, with silver preferring to adopt the +1 oxidation state, while copper being most commonly found at the +2 one. Simultaneously, there is immense difference of chemical reactivity of pristine $AgF_2$ and $CuF_2$, the former one being an extremely strong oxidizer, while the latter being not. Also, from the perspective of the electronic band structures these compounds are immensely different, with much more pronounced covalence of the metal-fluorine bonding for Ag than for Cu[19–21]. Finally, they both adopt related layered structures but there are differences in layer stacking between orthorhombic $AgF_2$ and its monoclinic Cu analogue (Fig.1). Indeed, $AgF_2$ may be considered as a distorted fluorite polytype, while $CuF_2$ is rutile-related[22,23]. Both $CuF_2$, $AgF_2$ and their derivatives attracted much attention as battery materials and as magnetic materials[5,14,24–28]. Our own efforts to prepare $Ag_{1-x}Cu_xF_2$ – aside from pure scientific curiosity – have been driven by two-decade long research where we targeted a possibility of electronic doping to $AgF_2$ and related systems[20,28–31]. As we will see, vast differences in the chemistry of copper and silver render this task quite difficult but fortunately not impossible.

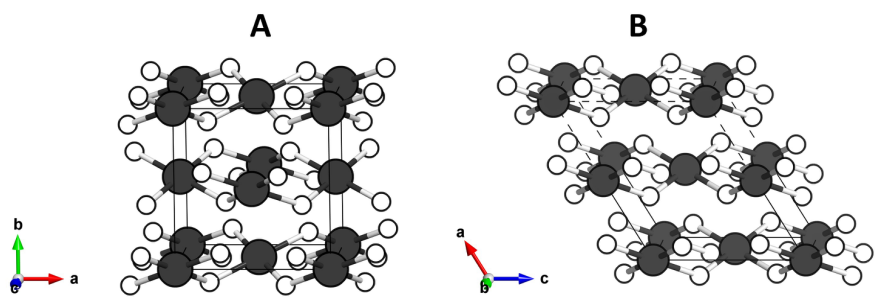

Fig.1 Crystal structure of $AgF_2$[32] (**A**) and $CuF_2$[33] (**B**). Color code: Ag, Cu – dark-grey, F – white. Only in-layer metal-fluorine bonds are shown. Solid lines – unit cells, dashed lines for supercell.

## 1. Results and discussion

### 1.1. Initial unsuccessful synthetic routes

Our initial attempts of preparation of $Ag_{1-x}Cu_xF_2$ have failed. They relied on two distinct procedures. The first one was heating of the mixture of $CuF_2$ and $AgF_2$ in argon-gas filled monel reactors at 400 °C for five hours. The second one used a similar procedure albeit at 300 °C, for 10 days, and in the presence of 1.5 bar $F_2$. The first approach yielded $Ag^{1+}Cu^{2+}F_3$[25] and AgF as the only products aside from initial reagents. The second approach gave no new phases at all. These results signified that in the presence of $CuF_2$ Lewis acid, $AgF_2$ becomes thermally unstable and it releases some $F_2$ gas, while the *in situ* formed reduced AgF may to some extent react via acid-base reaction yielding the previously known $AgCuF_3$. On the other hand, unsatisfying results of the second approach showed, that the presence of $F_2$ overpressure may prevent decomposition of $AgF_2$ but 300 °C is insufficient to produce conditions for ionic mobility and formation of the solid solution. Simultaneously, these conditions were aggressive enough to induce substantial corrosion of the monel container, thus increasing the risk of undesirable nickel contamination of the samples.

We have therefore attempted to prepare $Ag_{1-x}Cu_xF_2$ using the mechanochemical approach, as it proved successful for the preparation of other fluoride solid solutions[8,10,13]. Bearing in mind immense oxidizing properties of $AgF_2$, which must be used as a substrate of a mechanochemical reaction, we have utilized milling bowl equipped with milling disc, both made of pure nickel, and additionally equipped with cooling jacket. Even short attempts of milling of $AgF_2$ substrate alone with liquid nitrogen cooling have resulted in substantial corrosion of the milling vessel, yielding decomposition of initial substrate into AgF and other decomposition products – yet so far none of them were identified

(see ESI, section SI10). Since nickel milling vessel is not sufficiently chemically resistant towards $AgF_2$ during high-energy milling, further attempts to prepare $Ag_{1-x}Cu_xF_2$ with this method were suspended.

*1.2. Successful synthetic route*

In view of these failures, we have turned – without much enthusiasm – to conducting reactions inside a nickel reactor (130 ml) at much more aggressive conditions, i.e. at larger $F_2$ overpressure and simultaneously at much higher temperatures (Table 1). The detailed synthetic procedures are described below (cf. Methods and SI). We explored four molar ratios of $AgF_2/CuF_2$ with an increasing Cu content of 0.85/0.15 (**S1**), 0.70/0.30 (**S2**), and 1:1 (**S3**) annealed at around 530°C for 48 h, and one of 0.60:0.40 (**S4**); the latter was prepared at even higher temperature of 620°C. A typical starting $F_2$ overpressure was 4 bar. At such conditions, corrosion of the reactors is pronounced. Therefore, we have taken care to measure $F_2$ pressure after each reaction, and we have collected the specimen from the very middle of a nickel boat for analyses, while discarding the material adjacent to corroded nickel walls. Indeed, the subsequent EDX analyses indicated no nickel contamination in our samples. However, problems with leaks inevitably appeared at these reaction conditions; some reactors have become useless for new syntheses after applying these procedures and we were forced to drive conclusions from a rather limited number of attempts. Here, we report only the four cases, where the final $F_2$ overpressure after cooling the reactor to RT varied between 0.95 bar and 3.47 bar, suggesting no leak. We have subjected these samples to x-ray diffraction (XRD) measurements, and magnetic susceptibility analyses, as well as electronic impedance spectroscopy ones for selected samples. The XRD profiles were subsequently used for Rietveld analysis, where partial occupancy at metal sites was a free parameter.

*1.3. Results of the XRD analysis*

The chemical formulae for compounds based on Rietveld refinement of the XRD profiles for all samples are shown in Table 1.

Table 1. The chemical formulae for compounds based on Rietveld refinement of the XRD profiles. For detailed information check ESI (section SI3.)

| Samples | Compounds | | Starting ratio | The highest annealing point [°C] |
|---|---|---|---|---|
| | $AgF_2$-type $Pbca$ | $CuF_2$-type $P2_1/n$ | $AgF_2$:$CuF_2$ | |
| **S1** | $Ag_{0.85}Cu_{0.15}F_2$ | traces (single reflection) | 0.85:0.15 | 500 |
| **S2** | $Ag_{0.70}Cu_{0.30}F_2$ | traces (single reflection) | 0.70:0.30 | 530 |
| **S3** | $Ag_{0.71}Cu_{0.29}F_2$ | $Cu_{0.69}Ag_{0.31}F_2$ | 0.50:0.50 | 530 |
| **S4** | $Ag_{0.74}Cu_{0.26}F_2$ | $Cu_{0.56}Ag_{0.44}F_2$ | 0:60:0.40 | 620 |

The XRD profile of **S1** is extremely similar to that of pristine $AgF_2$, although reflections are shifted to higher angles, and relative intensities also differ as compared to binary $AgF_2$. The Rietveld fit shows that we have succeeded to obtain a polycrystalline sample of the $Pbca$ $AgF_2$-type structure with a partial Ag→Cu substitution, with traces from $CuF_2$-type phase, as indicated by the presence of a single weak reflection from the latter. The chemical formula from the unrestricted fit of occupancies (*cf.* ESI) of major phase is $Ag_{0.85}Cu_{0.15}F_2$ and thus corresponds ideally to the starting Ag/Cu molar ratio for this sample.

Sample **S2** was obtained in similar conditions to **S1**, but with the higher starting Cu contents. Here, we observed the formation of an analogous $AgF_2$–type orthorhombic phase, in which the cation substitution occurred to a greater degree of 30%, again in excellent agreement with the starting ratio of metal fluorides. In this case we also observed one additional reflection on XRD pattern with very low intensity. This feature may be tentatively assigned to a 011 reflection from a monoclinic $CuF_2$–type phase. This surmise was confirmed, as a greater amount of the same phase (albeit with somewhat different lattice constants) was present in the **S3**, for which starting molar ratio was 50%:50%. In **S3**, the new monoclinic phase is associated by the previously seen orthorhombic one ($AgF_2$–type). The chemical formula of $CuF_2$- and $AgF_2$-type phases in **S3** were $Cu_{0.69}Ag_{0.31}F_2$ and $Ag_{0.71}Cu_{0.29}F_2$, respectively (Table 1). This means, that despite of the higher initial Cu contents than for **S2**, there was no further enrichment in **S3** of the $AgF_2$-type phase in copper; instead, a separate phase of the $CuF_2$-type formed. This new phase has incorporated lots of silver, as well, and thus it corresponds to the second type of solid solution in this system. Importantly, the chemical composition of the mixture obtained from Rietveld refinement indicates the Ag:Cu molar ratio of 0.46:0:54, in reasonable agreement with the initial one of 0.50:0.50, thus testifying to sufficient accuracy of the Rietveld fit (Table SI3.2 in ESI).

Sample **S4** was also prepared with higher initial Cu content than **S2**, but this time we attempted to push miscibility even further by applying of a much higher temperature of annealing. Again, we observed the formation of two distinct phases (like in **S3**). However, we have not succeeded in the preparation of orthorhombic phase with the concentration level of Cu beyond 30%. The Ag:Cu ratio in $AgF_2$-type compound was refined to be 0.74:0.26, which is even slightly lower than for **S2** and **S3**. Nevertheless, the monoclinic $CuF_2$–type phase exhibited a significantly higher Ag content than in previous attempt (**S3**), with the molar ratio of Cu:Ag close to 0.56:0.44. This result shows that the orthorhombic phase has a limited capacity for Ag→Cu substitution, with the upper limit of Cu concentration of *ca.* 30%. Simultaneously, the CuF$_2$-type phase (which has additional angular degree of freedom, and lower symmetry than AgF$_2$-type) is more flexible, and Ag concentration may approach nearly 50% if temperature is sufficiently high.

To summarize, sample **S1** and **S2** are mainly Cu-poor phase with traces of the Cu-rich phase, while **S3** and **S4** resulted in phase separation of the initial starting ratio in Cu-poor and Cu-rich phases with different symmetry. But, are these genuine solid solutions? Note that, since Ag sublattice of AgF$_2$ is an orthorhombically distorted fcc one, one might expect that at least the 1:3, 3:1 or 1:1 stoichiometries could show some ordering on cationic sites. However, the fact that both AgF$_2$- and CuF$_2$-type phases may take diverse Ag:Cu ratios across a certain range is one argument for their genuine solid solution nature. Another argument comes from inspection of theoretical XRD profiles from ordered AgF$_2$-type Ag$_{0.75}$Cu$_{0.25}$F$_2$ phases, with three different types of ordering at metal sites (cf. ESI, section SI4). These analyses show the necessary appearance of new reflections for ordered phases, which are obviously absent in the disordered model; such reflections are absent in our XRD patterns, showing that **S1**–**S4** samples are disordered. Moreover, we do not see any departure from orthorhombicity, while ordered models usually have lower symmetry. All that is despite the fact that Ag$^{2+}$ and Cu$^{2+}$ have significantly different ionic radii (of 1.08 Å and 0.87 Å in octahedral environment, according to the Shannon-Prewitt definition[34] and thus a substantial mismatch of the all lattice constants, as well as a vast difference in ionicity of the metal-fluoride bonding. This result could be compared to those of our recent theoretical study[35], which has considered a possibility of the formation of a new Ag$^{(II)}$Cu$^{(II)}$F$_4$ compound. We have concluded that this ordered phase with two different sites for two types of metal cations would not be stable with respect to pristine binary fluorides at T → 0 K and p → 0 bar conditions, but it could be synthesized at elevated pressure of the several GPa, and possibly quenched down to ambient pressure conditions. Our own theoretical calculations for the ordered 3:1, 1:1 and 1:3 AgF$_2$- and CuF$_2$-type phases (cf. ESI) show that such substitutions are energetically disfavoured at T → 0 K and p → 0 bar conditions. However, the positive reaction enthalpies are rather small and of the order of 1–3 kJ/mol, which could be overcome by entropy factor at higher temperatures. It is unclear at the moment whether the disordered phases observed in this study are thermodynamically stable at ambient (p, T) conditions, but it is clear that at least at the conditions of their preparation they may form, and they do not phase separate in the end point phases x=0 or 1, during a slow quenching. That is, phase separation may occur but in Cu-rich and Cu-poor phases as obtained in **S3** and **S4** samples.

As we will see, an additional and simultaneously *the* key argument for the formation of genuine solid solutions will come, from magnetic studies (section 1.5).

*1.4. Deviation from the Vegard's law*

As already said, the two phases prepared inherit the orthorhombic and monoclinic cells of their parent (and closely related) AgF$_2$ and CuF$_2$ binaries. The formation of a solid solution rather than ordered types allows to relieve strain which is inevitably present if a system with separated and size- and volume-incompatible AgF$_2$ and CuF$_2$ sheets would form[35]. It is therefore interesting to check whether they follow the classical Vegard's law[36].

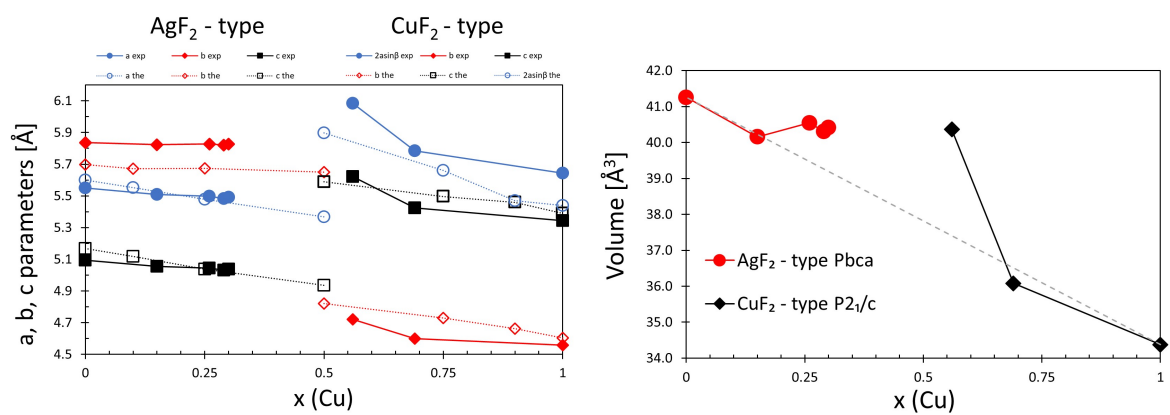

Fig. 2 Left panel: comparison between unit cell parameters from calculations (open symbols) and experimental data (filled symbols) vs doping level (x); For the monoclinic cell (CuF$_2$-type) we show twice the distance between planes given by 2a sin(β) (instead of a), as it can be directly compared with the parameter b of the orthorhombic structure having the same meaning (c.f. Fig.1). Computations were done using the VCA approximation without magnetism to simulate the disorder nature of the samples and obtain a smooth dependence of lattice parameters. *Cf.* ESI for detailed information. Right: volume of unit cell for Ag$_{1-x}$Cu$_x$F$_2$ vs doping (only experimental).

Fig. 2 presents comparison between experimental and theoretical (DFT, VCA) unit cell parameters and volumes for $Ag_{1-x}Cu_xF_2$ versus doping level (x). Based on Rietveld refinement, the lattice parameters and volume of $AgF_2$-type unit cells steadily decrease with the increasing Cu concentration. According to the experimental data, the 15% of copper doping in $AgF_2$ induces contraction of unit cell volume by 1.7%, comparing to pristine $AgF_2$. A larger level of substitution (ca. 30%) results a contraction by *ca.* 2.3% (**S2**) (see also ESI, Table SI3.1.). While a decrease of the unit cell volume is obviously expected, its extent is smaller than that predicted by the Vegard's law[36], (dashed line) and it also deviates from volume trend which has been theoretically predicted based on the virtual crystal approximation model without taking spin polarization into account[37] (see ESI, section SI6). Such a deviation from the Vegard's law suggests that the $AgF_2$ lattice is stiff enough to accommodate a moderate amount of $Cu^{2+}$ cations without substantially varying crystallographic unit cell parameters (Fig. 2). This is possible – and despite the substantial difference of ionic radii between $Ag^{2+}$ and $Cu^{2+}$ – because the Ag-F bonds are quite covalent[20,21], and [$AgF_2$] sheets are severely puckered, so they may react to incorporation of small $Cu^{2+}$ by simply flattening the sheet locally around the impurity site (note, bond angle change is always much less costly than bond shrinking or elongating). Indeed, DFT calculations show that Ag-F-Cu angles are on average 1-2 deg larger than the Ag-F-Ag angles in 3% Cu-doped $AgF_2$. Also, this type of structure has substantial flexibility in the direction perpendicular to the plane of propagation of the sheets, as the sheets are bound by rather weak secondary bonding. Hence, introduction of $Cu^{2+}$ instead of larger $Ag^{2+}$ does not induce much strain in this direction either. In consequence of that, $Cu^{2+}$ cations must have substantial local vibrational freedom ("rattling") when present at typical $Ag^{2+}$ sites.

Regarding the $CuF_2$-type structure, the volume and lattice parameters changes upon Cu→Ag substitution obviously follow opposite trend as that seen before for Ag→Cu substitution in the $AgF_2$-type structure, and the changes are much more pronounced than in the previous case (Fig. 2). Specifically, unit cell volume is larger by *ca.* 5.0%, compared to $CuF_2$ for $Cu_{0.69}Ag_{0.31}F_2$ and by *ca.* 17.4% larger for $Cu_{0.56}Ag_{0.44}F_2$. The largest strain is found in the direction perpendicular to the planes as it is the less bound direction (Fig. 3, see also ESI, table SI3.1.). Surprisingly, the smaller strain is found in the direction of the puckering despite being apparently more flexible than the other in-plane direction. The first composition does not fall far from the Vegard's law prediction, while the latter is placed nearby the trendline predicted from the VCA model lacking magnetism (Fig. 2). It is clear that introduction of larger $Ag^{2+}$ to a site of smaller $Cu^{2+}$ cation must have pronounced effect not only on a local coordination sphere, but also on overall lattice parameters. The strain introduced by such Cu→Ag substitution cannot be released solely by local puckering of the $CuF_2$ sheet but it also involves substantial changes to bond lengths and therefore to unit cell parameters. In other words, there is no "mirror symmetry" in Cu-doping to $AgF_2$ and Ag-doping to $CuF_2$! While analyzing these trends one should bear in mind that even the theoretically determined values using VCA may be intrinsically erroneous due to limitation of the method in the description of highly electron-correlated materials and doping effects[38]. Note that $AgF_2$ and $CuF_2$ are strongly correlated antiferromagnets with substantial magnetic superexchange[19–21,28], where any changes to chemical bonding result in changes of magnetic properties, and this impacts total energy of the system substantially; such effects are absent in diamagnets (say, in $ZnF_2$ with Zn→Cd substitution). Our simplistic DFT model does not grasp magnetic properties at all.

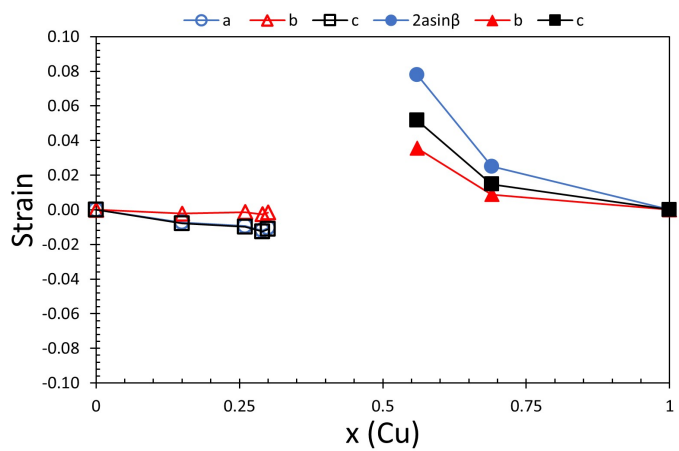

Fig. 3 Strain calculated for experimental unit cells (see ESI, section SI3) *vs.* x, *i.e.* Cu concentration. Right part corresponds to $CuF_2$ – type structures (filled symbols), left part shows strain for parameters for $AgF_2$ – type structures (open symbols). Strain of a given parameter for all structures were provided using following equation:

$$s = \frac{(p_x - p_0)}{p_0}$$

where: s – strain, $p_x$ – chosen parameter (a, b or c) of Cu-doped structure, $p_0$ – chosen parameter (a, b, or c) of pristine structure.

### 1.5. Magnetic properties

$AgF_2$ and $CuF_2$ order at, respectively, 163 K[33] and 69 K[33] with the corresponding intra-sheet magnetic superexchange constants of –70 meV[28] for the former and *ca.* –2.9 meV for the latter[39]. To the best of our knowledge, there is no Raman or neutron data that could yield accurate values of the magnetic interactions in $CuF_2$. However, accurate quantum chemistry computations predict a value of −11 meV[40]. The large difference in magnetic constant can be understand because the Cu-F bonds are more ionic than the Ag-F ones, and the former transmit the magnetic superexchange less effectively than the latter[41,42]. Since both $Ag^{2+}$ and $Cu^{2+}$ are paramagnetic s=½ cations, it is of immediate interest to study how the mutual substitutions in two lattices impact the magnetic properties of both phases. Such impact may be quite strong as documented by the $PdF_2$:Zn case[17], where the partial high-spin $Pd^{2+}$ with a diamagnetic $Zn^{2+}$ substitution at x=0.3 leads to reduction of the Néel point from 225 K to 9 K. From the theory side, it is known that for a nearest neighbor square lattice Heisenberg antiferromagnet, magnetic ordering vanishes for a concentration of non-magnetic impurities determined by the percolation threshold, $x_p$=0.407. While this has been experimentally verified in cuprates[43], the percolation threshold is very sensitive to the range of the magnetic interactions. For example, for a model with next and next-nearest neighbor interactions the critical concentration of non-magnetic impurities it shifts to $x_p$=0.592[44]. In the case of a solution of phases with different magnetic interactions as $AgF_2$ and $CuF_2$ one expects that $Ag_{1-x}Cu_xF_2$ has a crossover for x near $x_p$ from $AgF_2$-like to $CuF_2$-like magnetic properties.

Fig. 4 shows the $T_n$ (Néel temperature) changes *vs.* x value in $Ag_{1-x}Cu_xF_2$ systems together with those for the two binaries (end members of the series). For the **S1** sample for which the $CuF_2$-type component was not pronounced in XRD, we detected only the magnetic signals from the $AgF_2$-type phase, while for the remaining three samples we have detected signals from both polymorphs. As expected, the Néel temperature gets reduced with the increasing Cu contents. For $Ag_{0.85}Cu_{0.15}F_2$ (**S1**) the measured transition temperature is 146 K, while for the remaining three samples with x close to *ca.* 0.3, the $T_N$ goes down to *ca.* 136 K (**S2**, **S3**, **S4**). This tendency continues in the monoclinic phase. To understand this behavior, we have performed DFT+U calculations assuming several ordered structures and computed the average value of $|J_{2D}|$ assuming different magnetic configurations (see ESI for details).

According to our calculations for $Ag_{1-x}Cu_xF_2$, the progressive substitution of Ag by Cu, lowers the $|J_{2D}|$ value consistent with the experiments. For x = 0.25 the calculated average $|J2D|$ is by 31% lower as compared to pristine $AgF_2$ (*cf.* ESI, table SI5.1.). On the other hand, for $CuF_2$-type structure, the $|J_{2D}|$ value for x=0.75 is by 52% higher than for pristine $CuF_2$, and for equimolar mixing of cations in monoclinic $(Cu_{0.5}Ag_{0.5})F_2$ it is *ca.* 100% larger (hence, it doubles; *cf.* ESI, tables SI5.2. and SI9.2.).

The x=0.56 sample (**S4**, $Cu_{0.56}Ag_{0.44}F_2$, $CuF_2$-type) is particularly interesting. This sample is beyond the $x_p$=0.407 threshold for the nearest-neighbor model and therefore it should have $CuF_2$-like magnetic properties. Instead, its Néel temperature is close to the $AgF_2$-like samples. One possibility is that longer range interactions shift the relevant $x_p$ to larger values as mentioned above. Another possibility is that the Néel temperature is driven by the spin-direction anisotropy, and this is enhanced in the monoclinic structure. Indeed, a rough estimate of the Néel temperature using only the inter and intralayer interactions[45,46] in the end member compounds and neglecting the spin-direction anisotropy overestimates the Néel temperature of $AgF_2$ and underestimates the one $CuF_2$. This suggests that the spin-anisotropy contributes significantly to rise the Néel temperature of $CuF_2$ (while probably frustration effects depress the one of $AgF_2$). More theoretical and experimental work is needed to clarify this issue.

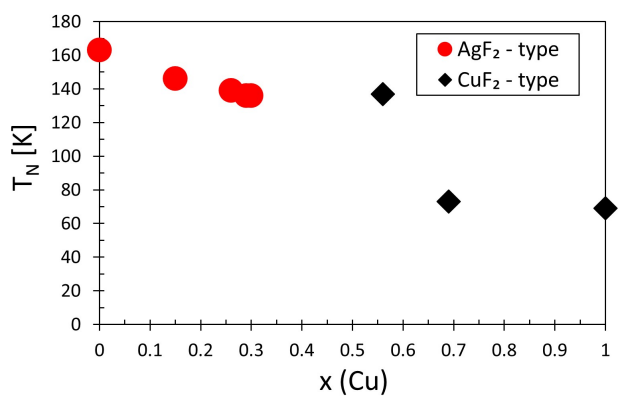

Fig. 4 $T_N$ changes *vs.* Cu concentration, x, in $AgF_2$– and $CuF_2$–type systems.

### 1.6. *Electric conductivity*

We have also examined the influence of cation mixing on the electric conductivity of the samples. We have focused our attention on the highest purity *i.e.* a single-phase sample $Ag_{0.85}Cu_{0.15}F_2$ (**S1**) due to the coexistence of two distinct phases in the remaining samples. $AgF_2$ and $CuF_2$ are rather poor ionic conductors[47], showing ionic conductivity at room temperature equal $1.85·10^{-13}$ S cm$^{−1}$ $2.64·10^{-15}$ S cm$^{−1}$ and, respectively (this work, *cf. E*SI, section SI8). Upon increasing of copper content in $Ag_{1-x}Cu_xF_2$ series, we observe a slight increase of specific ionic conductivity as compared to pristine $AgF_2$ (for x=0.15 of $3.71·10^{-13}$ S cm$^{−1}$, *i.e.* nearly twice as large as for parent compound). The fact that the changes are not dramatic indicates that synthetic procedures do not result in any substantial number of lattice defects such as vacancies or F adatoms (as that would increase ionic conductivity by orders of magnitude). In other words, we expect that the chemical composition of our samples is indeed very close to $MF_2$, regardless of the Cu/Ag ratio.

## Conclusions

The true solid solutions of transition metal fluorides with metallic elements originating from two different periods of the Periodic Table have so far been represented only by $PdF_2$:Zn. Here we have expanded this group by adding orthorhombic $AgF_2$:Cu and monoclinic $CuF_2$:Ag compositions to the pot. In the former, mutual solubility seems to be limited to x=0.3, while the latter may form a close-to-stoichiometric 1:1 solid solution while preserving its crystal structure. Ag→Cu and Cu→Ag substitutions impact magnetic properties of these phases to a great extent. Electric conductivity of a single-phase sample does not vary a lot with respect to pristine $AgF_2$, suggesting – regardless of the substitution at a metal site – the formation of a highly stoichiometric and defect-free $MF_2$ sample. It is interesting that such solid solutions form despite quite different chemical nature of $Ag^{(II)}$ and $Cu^{(II)}$, large discrepancies in the cationic radii, and the disparity of crystal structures of their binary fluorides. It is now tempting to expand this research even further, especially towards a multi-metal compositions, which could possibly constitute high-entropy systems. Whether such systems corresponding to genuine $MF_2$ stoichiometry could be prepared is unclear especially in view of the immense oxidizing power of $Ag^{2+}$ cation, and this is currently being tested in our laboratories.

## Acknowledgements

W.G. is grateful to the Polish National Science Center (NCN) for project Maestro (2017/26/A/ST5/00570). Research was carried out with the use of CePT infrastructure financed by the European Union – the European Regional Development Fund within the Operational Programme "Innovative economy" for 2007–2013 (POIG.02.02.00-14-024/08-00). ZM acknowledges the financial support of the Slovenian Research Agency (research core funding No. P1-0045; Inorganic Chemistry and Technology). Quantum mechanical calculations were performed using supercomputing resources of the ICM UW (project SAPPHIRE [GA83-34]).

## Methodology

**Computational details**

All calculations were performed within the DFT approach as implemented in VASP 5.4.4 code[48], using a GGA-type PBEsol[49] functional with projected augmented wave method[50,51]. The on-site Coulombic interactions of d electrons were realized by introducing Hubbard ($U_d$) and Hund ($J_H$) parameters (DFT+U formalism), as proposed by Liechtenstein[52]. $J_H$ was set to 1 eV[53], while the $U_d$ value was set to 8 and 10 eV, respectively for Ag and Cu[54,55]. Plane-wave cutoff energy of 520 eV was used in all systems. The k-spacing was set to 0.2 Å$^{-1}$ for geometry optimization and 0.14 Å$^{-1}$ for self-consistent-field convergence. The convergence thresholds of 10$^{-7}$ eV for electronic and 10$^{-5}$ ionic steps were used. Two possible paths of doping were theoretical considered. The first approach, ordered, based on the substitution of silver by copper, and vice versa, in $AgF_2$ and $CuF_2$ crystal structures. For small degree of doping, doped $AgF_2$ models were generated using 2x2x2 supercell, by substituting 1 or 2 Ag atoms with Cu (for 3.13 and 6.25% of doping, respectively, see SI). For 6.25% doping, we calculated 13 possible structures, with unique relative positions of two copper atoms in supercell. We chose the one with the lowest GS and describe in ESI. In calculations of doped $CuF_2$, we used 2x1x1 supercell or 1x1x1 cell (see SI). In disordered approach, the virtual crystal approximation (VCA) with weighted concentration of Cu and Ag atoms was used[37]. Magnetic models for ordered structures are presented in the ESI (section SI9).

**XRD measurements**

Powder samples were sealed under dry argon atmosphere inside 0.5 mm diameter perfluorinated quartz capillaries. Measurements were performed using PANalytical X'Pert Pro diffractometer, with a linear PIXcel Medipix2 detector (parallel beam of CoK$_{α1}$ and CoK$_{α2}$, radiation intensity ratio of 2:1), at room temperature. The parameters of $Ag_{1-x}Cu_xF$ and $Cu_{1-x}Ag_xF_2$ unit cells were refined by the Rietveld method in Jana2006 software[56], starting from the undoped $AgF_2$[32] and $CuF_2$[33], respectively, as a preliminary models. For peak shape description the Pseudo-Voight function was used, and Berar-Baldinozzi function to describe their asymmetry. The background was described by 20-36 Legendre polynomials. Symmetry restriction was set, with splitting of metal positions in unit cell for doped compounds. The overall sum of occupancies for atoms at the same positions were set to be equal 0.5, as the value for metal in undoped unit cell. The $U_{iso}$ parameter was set to be identical for both metal atoms in a shared position. For all analyses, we used absorption correction in the fashion provided by Jana2006 software, yielding positive $U_{iso}$ values for all atoms.

**Magnetic measurements**

15-40 mg of powder samples were loaded and closed inside teflon tubes under dry argon atmosphere; then transported into the measurement chamber. Magnetic measurements were performed using Quantum Design MPMS SQUID VSM magnetometer. The susceptibility vs temperature curves were measured between 2 and 300 K in different applied fields (FC and ZFC regimes) with 1K step. The obtained data has been corrected for the contribution of an empty sample holder.

**EIS measurements**

Impedance measurements were performed in the $10^{-2}$–$10^{6}$ Hz frequency range using a Solartron 1260 Frequency Response Analyzer with 1296 Dielectric Interface (see SI).

**Synthetic procedure**

The experiments are based on the solid-state reaction between different molar ratios of $AgF_2$ and $CuF_2$ compounds in $F_2$ atmosphere (see SI). The starting mixture was ground on agate mortar and sealed in a nickel reactor (130 ml) under dry argon atmosphere. The reactor was evacuated and then filled with 4 bar of fluorine gas. Under these conditions the different mixtures of (1-x) $AgF_2$ / x $CuF_2$ (with x = 0.15, 0.30, 0.40, and 0.50) were heated to about 500-530 °C (for x = 0.40 additionally to 620°C). After 24 h of heating, the reactor was cooled to room temperature, $F_2$ was pumped out, and the mixture was ground again. The synthesis procedure was repeated. After the second day of annealing, the dark brown or almost black samples were collected and analyzed. The samples were taken from the middle part of the reactor. Possible contamination by nickel (from the reactor) was excluded on the basis of results of the EDX analysis.

All figures of structures were visualized using VESTA software[57].

# Bibliography

[1] G. Perrault, *The Canadian Mineralogist* **1968**, *9*, 383–402.
[2] J.-W. Yeh, S.-K. Chen, S.-J. Lin, J.-Y. Gan, T.-S. Chin, T.-T. Shun, C.-H. Tsau, S.-Y. Chang, *Advanced Engineering Materials* **2004**, *6*, 299–303.
[3] C. M. Rost, E. Sachet, T. Borman, A. Moballegh, E. C. Dickey, D. Hou, J. L. Jones, S. Curtarolo, J.-P. Maria, *Nat Commun* **2015**, *6*, 8485.
[4] T. Wang, H. Chen, Z. Yang, J. Liang, S. Dai, *J. Am. Chem. Soc.* **2020**, *142*, 4550–4554.
[5] X. Hua, R. Robert, L.-S. Du, K. M. Wiaderek, M. Leskes, K. W. Chapman, P. J. Chupas, C. P. Grey, *J. Phys. Chem. C* **2014**, *118*, 15169–15184.
[6] M. G. Cottam, D. J. Lockwood, *Low Temperature Physics* **2019**, *45*, 78–91.
[7] F. Waltz, M. A. Swider, P. Hoyer, T. Hassel, M. Erne, K. Möhwald, M. Adlung, A. Feldhoff, C. Wickleder, F.-W. Bach, P. Behrens, *J Mater Sci* **2012**, *47*, 176–183.
[8] G. Scholz, S. Breitfeld, T. Krahl, A. Düvel, P. Heitjans, E. Kemnitz, *Solid State Sciences* **2015**, *50*, 32–41.
[9] X. Mu, W. Sigle, A. Bach, D. Fischer, M. Jansen, P. A. van Aken, *Z. anorg. allg. Chem.* **2014**, *640*, 1868–1875.
[10] M. Heise, G. Scholz, A. Düvel, P. Heitjans, E. Kemnitz, *Solid State Sciences* **2016**, *60*, 65–74.
[11] A. Düvel, P. Heitjans, P. P. Fedorov, V. V. Voronov, A. A. Pynenkov, K. N. Nishchev, *Solid State Sciences* **2018**, *83*, 188–191.
[12] $Ca_{1-x}Y_xF_{2+x}$ exemplifies a very interesting system with a highly increased ionic mobility as compared to its binary counterparts, but it cannot be considered as a genuine solid solution, since there are $F_x$ ad-anions in its structure. P. P. Fedorov, O. E. Izotova, V. B. Alexandrov, B. P. Sobolev, *Journal of Solid State Chemistry* **1974**, *9*, 368–374.
[13] P. Anitha Sukkurji, Y. Cui, S. Lee, K. Wang, R. Azmi, A. Sarkar, S. Indris, S. S. Bhattacharya, R. Kruk, H. Hahn, Q. Wang, M. Botros, B. Breitung, *Journal of Materials Chemistry A* **2021**, *9*, 8998–9009.
[14] D. T. Thieu, M. H. Fawey, H. Bhatia, T. Diemant, V. S. K. Chakravadhanula, R. J. Behm, C. Kübel, M. Fichtner, *Advanced Functional Materials* **2017**, *27*, 1701051.
[15] C. Villa, S. Kim, Y. Lu, V. P. Dravid, J. Wu, *ACS Appl. Mater. Interfaces* **2019**, *11*, 647–654.
[16] F. Wang, S.-W. Kim, D.-H. Seo, K. Kang, L. Wang, D. Su, J. J. Vajo, J. Wang, J. Graetz, *Nat Commun* **2015**, *6*, 6668.
[17] D. Paus, R. Hoppe, *Z. Anorg. Allg. Chem.* **1977**, *431*, 207–216.
[18] R. Schmidt, M. Kraus, B. G. Müller, *Z. anorg. allg. Chem.* **2001**, *627*, 2344–2350.
[19] C. Miller, A. S. Botana, *Phys. Rev. B* **2020**, *101*, 195116.
[20] W. Grochala, R. Hoffmann, *Angew. Chem. Int. Ed.* **2001**, *40*, 2742–2781.
[21] W. Grochala, R. G. Egdell, P. P. Edwards, Z. Mazej, B. Žemva, *ChemPhysChem* **2003**, *4*, 997–1001.
[22] A. Grzelak, J. Gawraczyński, T. Jaroń, D. Kurzydłowski, A. Budzianowski, Z. Mazej, P. J. Leszczyński, V. B. Prakapenka, M. Derzsi, V. V. Struzhkin, W. Grochala, *Inorg. Chem.* **2017**, *56*, 14651–14661.
[23] A. Grzelak, J. Gawraczyński, T. Jaroń, D. Kurzydłowski, Z. Mazej, P. J. Leszczyński, V. B. Prakapenka, M. Derzsi, V. V. Struzhkin, W. Grochala, *Dalton Trans.* **2017**, *46*, 14742–14745.
[24] J. Tong, C. Lee, M.-H. Whangbo, R. K. Kremer, A. Simon, J. Köhler, *Solid State Sciences* **2010**, *12*, 680–684.
[25] W. Tong, G. G. Amatucci, *Journal of Power Sources* **2017**, *362*, 86–91.
[26] G. G. Amatucci, N. Pereira, *Journal of Fluorine Chemistry* **2007**, *128*, 243–262.
[27] W. Tong, W.-S. Yoon, N. M. Hagh, G. G. Amatucci, *Chem. Mater.* **2009**, *21*, 2139–2148.
[28] J. Gawraczyński, D. Kurzydłowski, R. A. Ewings, S. Bandaru, W. Gadomski, Z. Mazej, G. Ruani, I. Bergenti, T. Jaroń, A. Ozarowski, S. Hill, P. J. Leszczyński, K. Tokár, M. Derzsi, P. Barone, K. Wohlfeld, J. Lorenzana, W. Grochala, *Proc Natl Acad Sci USA* **2019**, *116*, 1495–1500.
[29] W. Grochala, *J Supercond Nov Magn* **2018**, *31*, 737–752.
[30] D. Jezierski, A. Grzelak, X. Liu, S. K. Pandey, M. N. Gastiasoro, J. Lorenzana, J. Feng, W. Grochala, *Phys. Chem. Chem. Phys.* **2022**, *24*, 15705–15717.
[31] A. Grzelak, J. Lorenzana, W. Grochala, *Angew. Chem. Int. Ed.* **2021**, *60*, 13892–13895.

[32] P. Charpin, P. Plurien, P. Mériel, *Bulletin de Minéralogie* **1970**, *93*, 7–13.
[33] P. Fischer, W. Hälg, D. Schwarzenbach, H. Gamsjäger, *Journal of Physics and Chemistry of Solids* **1974**, *35*, 1683–1689.
[34] R. D. Shannon, C. T. Prewitt, *Acta Crystallogr B Struct Sci* **1969**, *25*, 925–946.
[35] M. A. Domański, M. Derzsi, W. Grochala, *RSC Adv.* **2021**, *11*, 25801–25810.
[36] L. Vegard, *Z. Physik* **1921**, *5*, 17–26.
[37] L. Bellaiche, D. Vanderbilt, *Phys. Rev. B* **2000**, *61*, 7877–7882.
[38] C. Eckhardt, K. Hummer, G. Kresse, *Phys. Rev. B* **2014**, *89*, 165201.
[39] R. J. Joenk, R. M. Bozorth, *Journal of Applied Physics* **1965**, *36*, 1167–1168.
[40] P. Reinhardt, I. de P. R. Moreira, C. de Graaf, R. Dovesi, F. Illas, *Chemical Physics Letters* **2000**, *319*, 625–630.
[41] J. B. Goodenough, *Phys. Rev.* **1955**, *100*, 564–573.
[42] J. Kanamori, *Journal of Physics and Chemistry of Solids* **1959**, *10*, 87–98.
[43] O. P. Vajk, P. K. Mang, M. Greven, P. M. Gehring, J. W. Lynn, *Science* **2002**, *295*, 1691–1695.
[44] K. Malarz, S. Galam, *Phys. Rev. E* **2005**, *71*, 016125.
[45] C. Yasuda, S. Todo, K. Hukushima, F. Alet, M. Keller, M. Troyer, H. Takayama, *Phys. Rev. Lett.* **2005**, *94*, 217201.
[46] S. J. Blundell, *Contemporary Physics* **2007**, *48*, 275–290.
[47] W. Adamczyk, P. Połczyński, A. Mika, T. Jaroń, Z. Mazej, K. J. Fijalkowski, R. Jurczakowski, W. Grochala, *Journal of Fluorine Chemistry* **2015**, *174*, 22–29.
[48] G. Kresse, J. Furthmüller, *Phys. Rev. B* **1996**, *54*, 11169–11186.
[49] J. P. Perdew, A. Ruzsinszky, G. I. Csonka, O. A. Vydrov, G. E. Scuseria, L. A. Constantin, X. Zhou, K. Burke, *Phys. Rev. Lett.* **2008**, *100*, 136406.
[50] P. E. Blöchl, *Phys. Rev. B* **1994**, *50*, 17953–17979.
[51] G. Kresse, D. Joubert, *Phys. Rev. B* **1999**, *59*, 1758–1775.
[52] A. I. Liechtenstein, V. I. Anisimov, J. Zaanen, *Phys. Rev. B* **1995**, *52*, R5467–R5470.
[53] V. I. Anisimov, J. Zaanen, O. K. Andersen, *Phys. Rev. B* **1991**, *44*, 943–954.
[54] R. Piombo, D. Jezierski, H. P. Martins, T. Jaroń, M. N. Gastiasoro, P. Barone, K. Tokár, P. Piekarz, M. Derzsi, Z. Mazej, M. Abbate, W. Grochala, J. Lorenzana, *Phys. Rev. B* **2022**, *106*, 035142.
[55] A. B. Kunz, *Phys. Rev. B* **1982**, *26*, 2070–2075.
[56] V. Petříček, M. Dušek, L. Palatinus, *Zeitschrift für Kristallographie - Crystalline Materials* **2014**, *229*, 345–352.
[57] K. Momma, F. Izumi, *J Appl Cryst* **2008**, *41*, 653–658.

# SUPPORTING INFORMATION

# Unexpected coexisting solid solutions in the quasi-binary Ag(II)F$_2$/Cu(II)F$_2$ phase diagram

D. Jezierski,[a] K. Koteras,[a] M. Domański, [a] P. Połczyński,[b] Z. Mazej,[c] J. Lorenzana,[d] W. Grochala*[a]

[a] Centre of New Technologies, University of Warsaw, 02097 Warsaw, Poland
[b] Faculty of Chemistry, Biological and Chemical Research Centre, University of Warsaw, 02089 Warsaw, Poland
[c] Department of Inorganic Chemistry and Technology, Jožef Stefan Institute, 1000 Ljubljana, Slovenia
[d] Institute for Complex Systems (ISC), Consiglio Nazionale delle Ricerche, Dipartimento di Fisica, Università di Roma "La Sapienza", 00185 Rome, Italy

* w.grochala@cent.uw.edu.pl

## List of contents

# SI1. Detailed synthetic procedures

**Table SI1.1.** *Detailed synthetic procedures of successful attempts.*

| Sample | Starting material | p($F_2$) / bar | appearance of powder | heating procedure / °C | p($F_2$) after synth. / bar |
|---|---|---|---|---|---|
| **S1** | 0.15$CuF_2$+0.85$AgF_2$ (1.08 mmol + 6.12 mmol) | 4 | dark brown | I. 500 (24h) II. 440 (24h) | I. ~1.92 II. ~3.47 |
| **S2** | 0.30$CuF_2$+0.70$AgF_2$ (0.91 mmol + 2.11 mmol) | 4 | dark brown | I. 530 °C (24 h) II. 530 °C (24 h) | I. ~3.47 II. ~3.47 |
| **S3** | 0.50$CuF_2$+0.50$AgF_2$ (1.62 mmol + 1.62 mmol) | 4 | dark brown | I. 530 °C (24h) II. 530 °C (24h) | I. ~3.04 II. ~3.04 |
| **S4** | 0.40$CuF_2$+0.60$AgF_2$ (7.81 mmol + 11.71 mmol) | 4 | black | I. 500 (24h) II. 623 (24h) | I. ~1.74 II. ~0.95 |

# SI2. Experimental XRD patterns

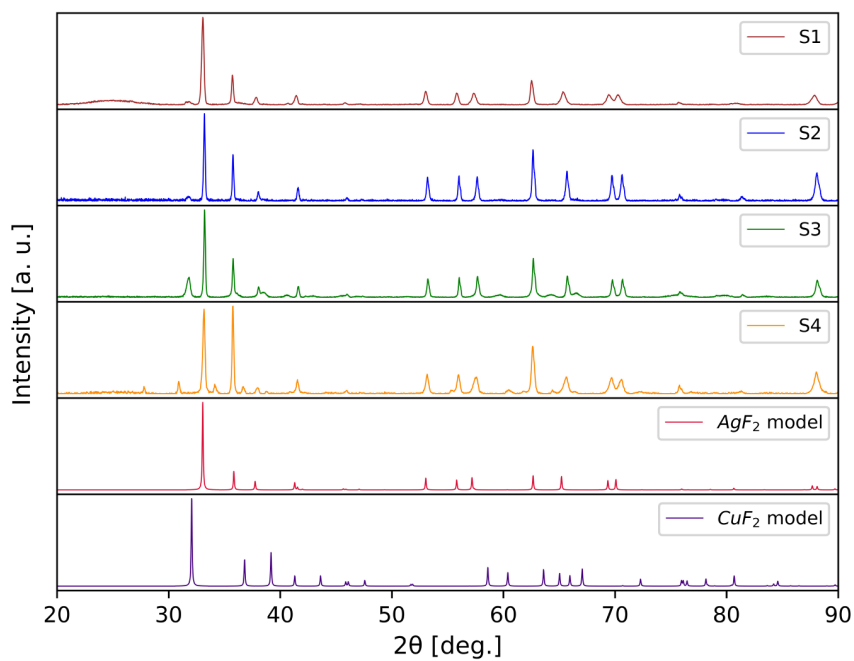

**Figure SI2.1.** XRD patterns for samples **S1-S4** and binary $AgF_2$ and $CuF_2$ references. Co beam was used.

# SI3. Crystallographic unit cell parameters and results of Rietveld refinement

**Table SI3.1.** *Cell parameters (a, b, c, V) after Rietveld refinement and the Rietveld refinement parameters. The columns "%d" corresponds to percentage difference between parameter for the pristine compound and the related solid solution ($AgF_2$- or $CuF_2$-type).*

| Sample | Compound | Parameters | | | | | | | | | Rietveld refinement | | |
|---|---|---|---|---|---|---|---|---|---|---|---|---|---|
| | | a | %d | b | %d | c | %d | V | %d | beta | GoF | Rp | wRp |
| ---- | $AgF_2$ | 5.550 | | 5.836 | | 5.095 | | 41.257 | | | 2.03 | 2.04 | 3.17 |
| | $CuF_2$ | 3.294 | | 4.559 | | 5.345 | | 34.377 | | 121.041 | 4.23 | 3.60 | 5.74 |
| *S1* | $Ag_{0.85}Cu_{0.15}F_2$ | 5.510 | -0.73 | 5.823 | -0.22 | 5.055 | -0.78 | 40.545 | -1.72 | | 1.76 | 1.33 | 2.09 |
| *S2* | $Ag_{0.70}Cu_{0.30}F_2$ | 5.492 | -1.04 | 5.827 | -0.15 | 5.038 | -1.11 | 40.312 | -2.29 | | 1.26 | 1.40 | 1.90 |
| *S3* | $Ag_{0.71}Cu_{0.29}F_2$ | 5.484 | -1.19 | 5.821 | -0.26 | 5.031 | -1.25 | 40.155 | -2.67 | | 1.06 | 1.17 | 1.57 |
| | $Cu_{0.69}Ag_{0.31}F_2$ | 3.386 | +2.80 | 4.599 | +0.88 | 5.424 | +1.49 | 36.082 | +4.96 | 121.312 | | | |
| *S4* | $Ag_{0.74}Cu_{0.26}F_2$ | 5.498 | -0.93 | 5.828 | -0.13 | 5.045 | -0.99 | 40.412 | -2.05 | | 1.61 | 1.34 | 2.11 |
| | $Cu_{0.56}Ag_{0.44}F_2$ | 3.588 | +8.95 | 4.721 | 3.55 | 5.621 | 5.17 | 40.365 | +17.42 | 122.019 | | | |

**Table SI3.2** *Detailed data from Rietveld refinement. **U**$_{iso}$ is mean square displacement amplitude of the atom averaged over all directions for **cation** (Cu and/or Ag) and **F** – fluorine.  **A**$_i$ corresponds to the occupancy of Ag or Cu of the crystallographic site.*

| Sample | Compound | U$_{iso}$ | | a$_i$ (occupancy) | | Metal ratio | | Relative mass | uncertainty [%] | AgF$_2$:CuF$_2$ ratio (Rietveld) |
|---|---|---|---|---|---|---|---|---|---|---|
| | | Cation | F | Ag | Cu | Ag | Cu | | | |
| **S1** | Ag$_{0.85}$Cu$_{0.15}$F$_2$ | 0.009 | 0.019 | 0.43 | 0.07 | 0.85 | 0.15 | 1 | --- | 0.85:0:15 |
| **S2** | Ag$_{0.70}$Cu$_{0.30}$F$_2$ | 0.012 | 0.028 | 0.35 | 0.15 | 0.7 | 0.3 | 1 | --- | 0.70:0:30 |
| **S3** | Ag$_{0.71}$Cu$_{0.29}$F$_2$ | 0.083 | 0.120 | 0.35 | 0.15 | 0.71 | 0.29 | 0.60 | 2 | 0.46:0:54 |
| | Cu$_{0.69}$Ag$_{0.31}$F$_2$ | 0.067 | 0.050 | 0.16 | 0.34 | 0.31 | 0.69 | 0.40 | 18 | |
| **S4** | Ag$_{0.74}$Cu$_{0.26}$F$_2$ | 0.033 | 0.052 | 0.37 | 0.13 | 0.74 | 0.26 | 0.756 | 10 | 0.62:0.38 |
| | Cu$_{0.56}$Ag$_{0.44}$F$_2$ | 0.030 | 0.066 | 0.22 | 0.28 | 0.44 | 0.56 | 0.244 | 9 | |

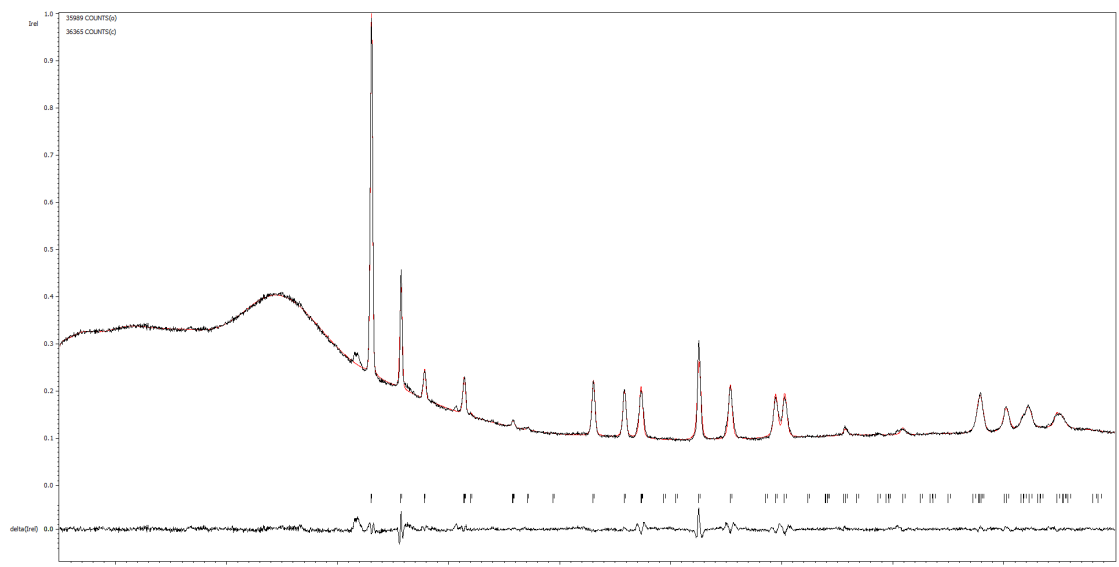

**Fig. SI3.1.** Rietveld refinement for **S1**. Red line corresponds to model pattern for refinement Ag$_{0.85}$Cu$_{0.15}$F$_2$ structure.

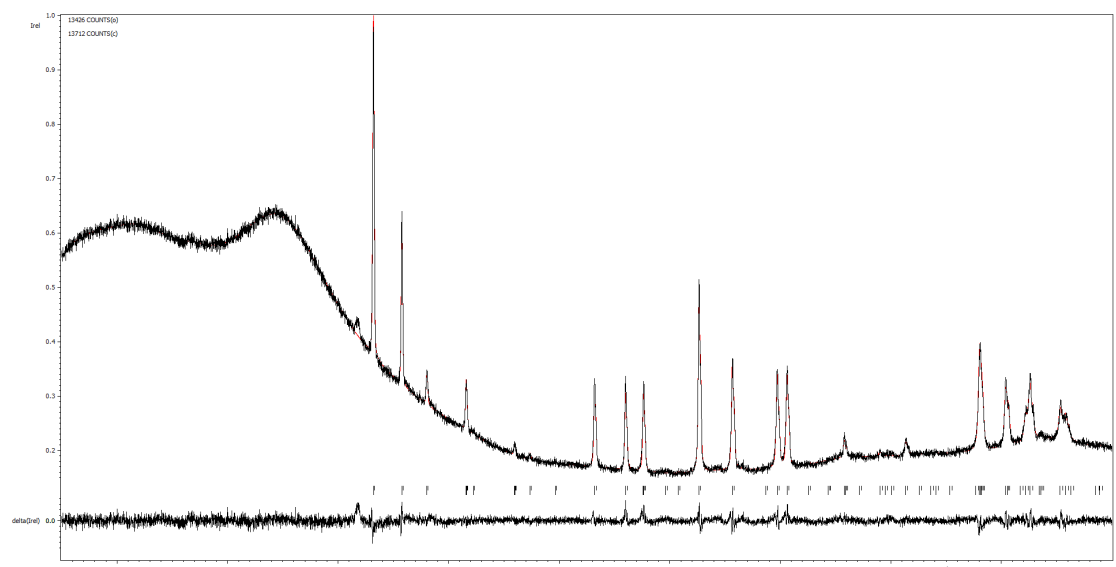

**Fig. SI3.2.** Rietveld refinement for **S2**. Red line corresponds to model pattern for refinement Ag$_{0.70}$Cu$_{0.30}$F$_2$ structure.

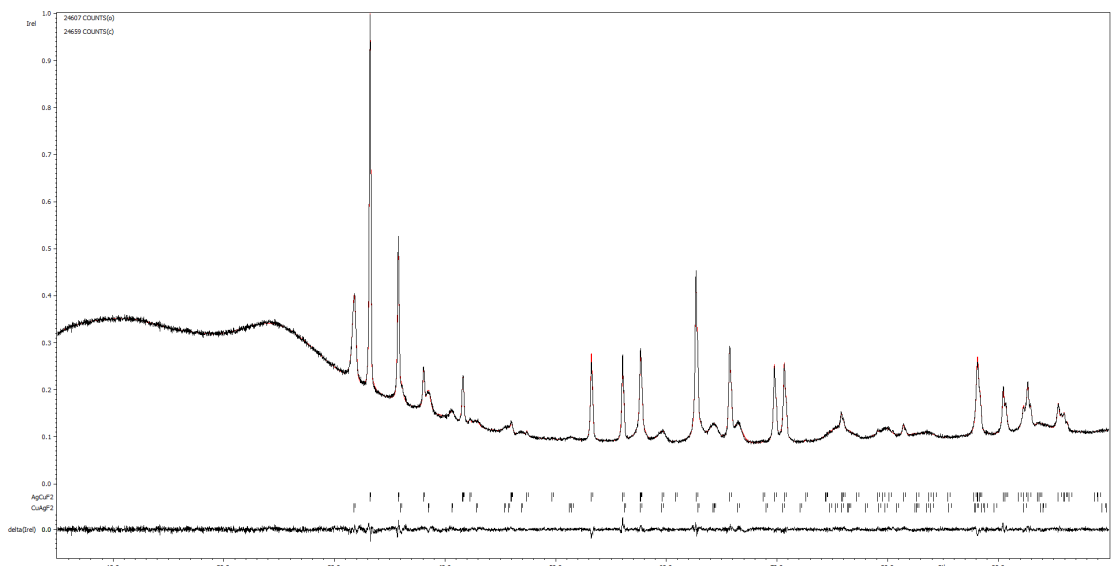

**Fig. SI3.3.** Rietveld refinement for **S3**. Red line corresponds to model pattern for refinement $Cu_{0.69}Ag_{0.31}F_2$ and $Ag_{0.71}Cu_{0.29}F_2$ structures.

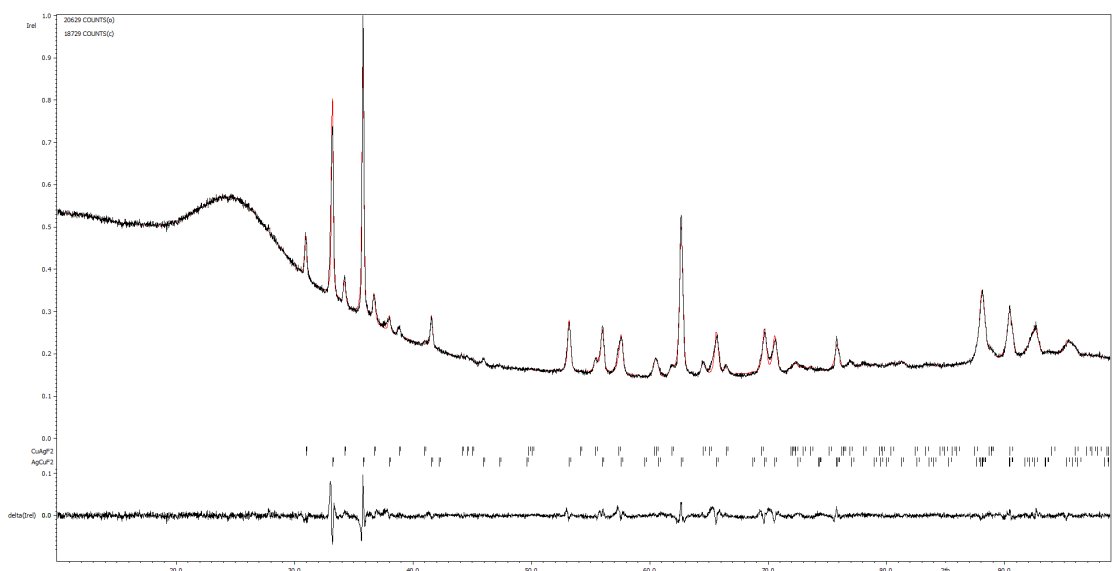

**Fig. SI3.4.** Rietveld refinement for **S4**. Red line corresponds to model pattern for refinement $Ag_{0.74}Cu_{0.26}F_2$ and $Cu_{0.56}Ag_{0.44}F_2$ structures.

## SI4. Theoretical XRD patterns for diverse ordered Ag$_{0.75}$Cu$_{0.25}$F$_2$ phases

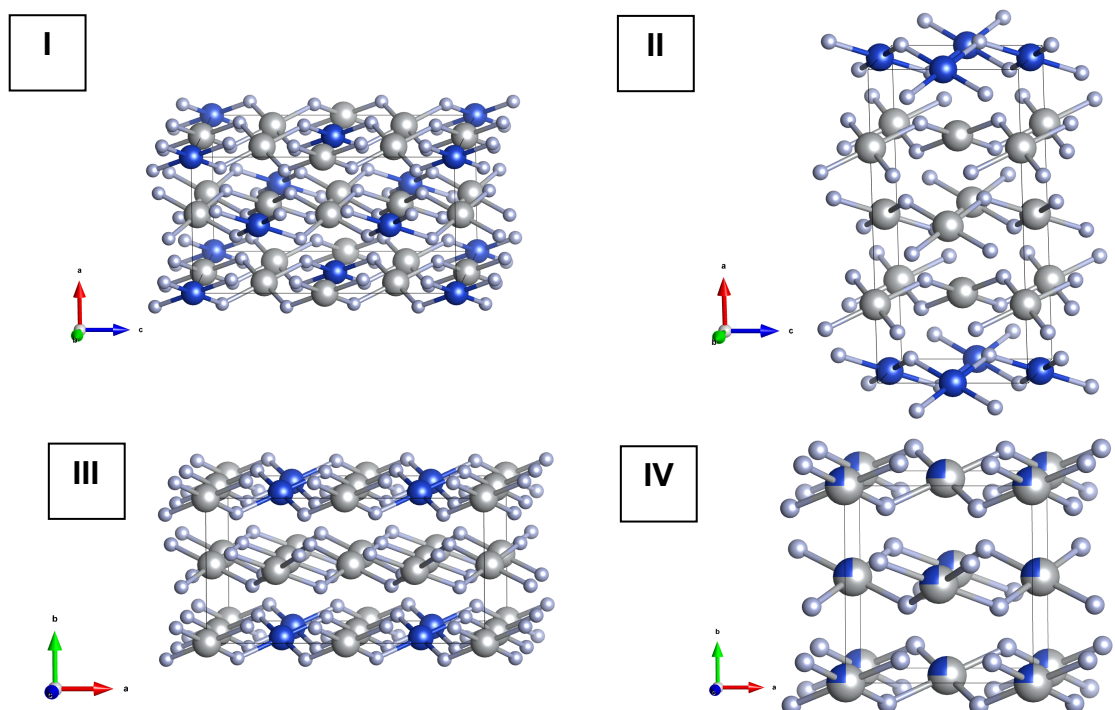

**Fig. SI4.1.** Ordered models **I-III** and **IV** disordered corresponding to Ag$_{0.75}$Cu$_{0.25}$F$_2$

All unit cell parameters for these models were set identical to those from Rietveld refinement for Ag$_{0.70}$Cu$_{0.30}$F$_2$ (**S3**, **Table SI3.2.**)

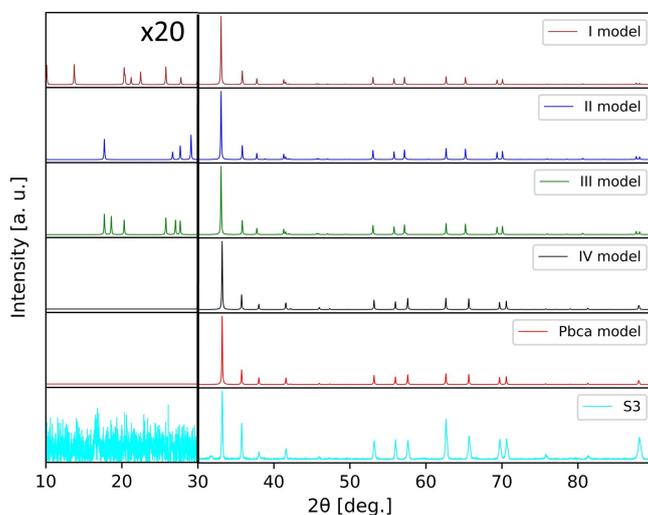

**Fig. SI4.2**. Comparison XRD patterns between ordered **I-III** and disordered **IV** structures from **Fig. SI5.** Additionally, AgF$_2$ model (*Pbca*) and **S3** pattern was added. The intensities of peaks for 10-30 2theta part were magnified 20 times to show that ordered phases are not seen in experimental profile.

# SI5. DFT+U calculations for AgF$_2$/CuF$_2$ phases with ordered metal sites

**Table SI5.1.** *Parameters of AgF$_2$ cells in case of Cu doping. Based on the DFT+U calculations. U$_{Ag}$ = 8eV, U$_{Cu}$ = 10eV, J for Ag and Cu equal to 1eV. GREEN rows: 2x2x2 supercells calculations. "%diff" is a percentage difference of specific parameter referring to pristine AgF$_2$ (DFT+U). The calculations indicated antiferromagnetic ordering as a ground state magnetic ordering for all structures. The columns "%d" has the same meanings as in **Table SI3.1**.*

| Doping [%] | Formula | a [Å] | % diff | b [Å] | % diff | c [Å] | % diff | α [°] | β [°] | γ [°] | V [Å$^3$] | V diff [%] | J$_{2D}$ [meV]* | dE [kJ/mol]** | dV$^\&$ | T eq. [K] # |
|---|---|---|---|---|---|---|---|---|---|---|---|---|---|---|---|---|
| 0 | AgF$_2$ | 5.433 | | 5.008 | | 5.769 | | 90.000 | 90.000 | 90.000 | 39.245 | | -32.70 | --- | --- | --- |
| 3.13 | Ag$_{0.97}$Cu$_{0.03}$F$_2$ | 5.413 | -0.37 | 4.985 | -0.47 | 5.766 | -0.05 | 89.927 | 90.552 | 90.008 | 38.866 | -0.96 | --- | +0.63 | -0.17 | --- |
| 6.25 | Ag$_{0.94}$Cu$_{0.06}$F$_2$ | 5.401 | -0.59 | 4.972 | -0.73 | 5.754 | -0.26 | 89.817 | 89.108 | 89.987 | 38.622 | -1.59 | --- | +1.26 | -0.20 | --- |
| 25 | Ag$_{0.75}$Cu$_{0.25}$F$_2$ | 5.323 | -2.02 | 4.909 | -1.98 | 5.683 | -1.48 | 89.537 | 94.189 | 89.971 | 37.029 | -5.65 | -22.49 | +3.12 | -0.53 | --- |
| 50 | Ag$_{0.50}$Cu$_{0.50}$F$_2$ | 5.416 | -0.32 | 4.718 | -5.80 | 5.948 | +3.11 | 89.960 | 105.949 | 90.009 | 36.534 | -6.91 | -18.62 | +2.24 | +0.65 | 546 |
| 75 | Ag$_{0.25}$Cu$_{0.75}$F$_2$ | 5.367 | -1.22 | 4.360 | -12.94 | 5.924 | +2.69 | 90.213 | 108.455 | 89.832 | 34.715 | -11.54 | -14.05 | +2.38 | +0.52 | 1384 |
| 100 | CuF$_2$: P2$_1$/c | 5.294 | -2.55 | 4.501 | -10.14 | 5.801 | +0.56 | 89.999 | 109.944 | 89.999 | 32.485 | -17.22 | -9.15 | --- | --- | --- |
| 100 | CuF$_2$: Pbca | 4.971 | -8.50 | 4.668 | -6.80 | 5.310 | -7.96 | 90.000 | 90.000 | 90.000 | 30.803 | -21.51 | -6.61 | --- | --- | --- |

Note, the 50:50 phase (Ag$_{0.50}$Cu$_{0.50}$F$_2$) is identical to that described in Ref [2] and it is also, according to the DFT+U calculations, the most stable ordered polytype for this stoichiometry. The J$_{2D}$ for this structure is calculated as an average of superexchange constants in two types of planes Ag-F-Ag and Cu-F-Cu.

**Table SI5.2.** *Parameters of CuF$_2$ cells in case of Ag doping. Based on the DFT+U calculations. U$_{Ag}$ = 8eV, U$_{Cu}$ = 10eV, J for Ag and Cu equal to 1eV. The "%diff" column is a percentage difference of specific parameter referring to pristine CuF$_2$ (DFT+U). The calculations indicated antiferromagnetic ordering as a ground state magnetic ordering for all structures.*

| Cell | Doping [%] | Formula | a [Å] | % diff | b [Å] | % diff | c [Å] | % diff | α [°] | β [°] | γ [°] | V [Å$^3$] | V diff [%] | J$_{2D}$ [meV]* | dE [kJ/mol]** | dV$^\&$ | T eq. [K] # |
|---|---|---|---|---|---|---|---|---|---|---|---|---|---|---|---|---|---|
| 1x1x1 | 0 | CuF$_2$ | 3.192 | | 4.508 | | 5.291 | | 90.000 | 121.328 | 90.000 | 32.518 | | -8.27 | --- | --- | --- |
| 1x1x1 | 50 | Cu$_{0.50}$Ag$_{0.50}$F$_2$ | 3.383 | +5.98 | 4.672 | +3.65 | 5.485 | +3.68 | 90.175 | 105.949 | 89.823 | 37.340 | +14.83 | -20.07 | +3.91 | +1.46 | 1523 |
| 1x1x1 | 100 | AgF$_2$: P2$_1$/c | 3.636 | +13.89 | 4.775 | +5.92 | 5.706 | +7.84 | 90.000 | 122.110 | 90.000 | 41.949 | +29.00 | -40.09 | --- | --- | --- |
| 2x1x1 | 0 | CuF$_2$ | 6.384 | | 4.506 | | 5.294 | | 90.000 | 121.325 | 90.000 | 32.521 | | -8.22 | --- | --- | --- |
| 2x1x1 | 25 | Cu$_{0.75}$Ag$_{0.25}$F$_2$ | 6.593 | +3.27 | 4.599 | +2.08 | 5.365 | +1.34 | 90.157 | 121.748 | 90.024 | 34.587 | +6.36 | -12.51 | +2.57 | +0.39 | 3790 |
| 2x1x1 | 50 | Cu$_{0.50}$Ag$_{0.50}$F$_2$ | 6.786 | +6.28 | 4.685 | +3.98 | 5.429 | +2.54 | 90.000 | 122.580 | 90.000 | 36.353 | +11.80 | -16.38 | +2.59 | +0.47 | 3129 |
| 2x1x1 | 75 | Cu$_{0.25}$Ag$_{0.75}$F$_2$ | 6.920 | +8.39 | 4.740 | +5.20 | 5.562 | +5.07 | 89.790 | 122.102 | 90.064 | 38.636 | +18.81 | -27.68 | +2.85 | +1.07 | 1509 |
| 2x1x1 | 100 | AgF$_2$: P2$_1$/c | 7.258 | +13.68 | 4.761 | +5.66 | 5.682 | +7.34 | 90.003 | 121.643 | 90.017 | 41.784 | +28.50 | -37.98 | --- | --- | --- |

*J = E$_{AFM}$ - E$_{FM}$, corresponding to in-layer superexchange; ** dE = E$_{Ag_xCu_yF_2}$ - (xE$_{AgF_2}$ + yE$_{CuF_2}$); $^\&$ dV = V$_{Ag_xCu_yF_2}$ - (xV$_{AgF_2}$ + yV$_{CuF_2}$); # T$_{eq}$ = $\frac{dE}{dS}$, where dS is related to dV, dS/dV = 1.76 Jmol$^{-1}$K$^{-1}$[1]

The 2x1x1 Cu$_{0.50}$Ag$_{0.50}$F$_2$ consists of alternating [AgF$_4$]$^{2-}$ and [CuF$_4$]$^{2-}$ planes – the most stable ordered polytype for this stoichiometry. For this structure, the J$_{2D}$ is an average of superexchange constants in two types of planes. Determination of J values for disordered Cu$_{0.50}$Ag$_{0.50}$F$_2$ structure with Ag-F-Cu layers is presented in section **SI9** of this work (below, ESI).

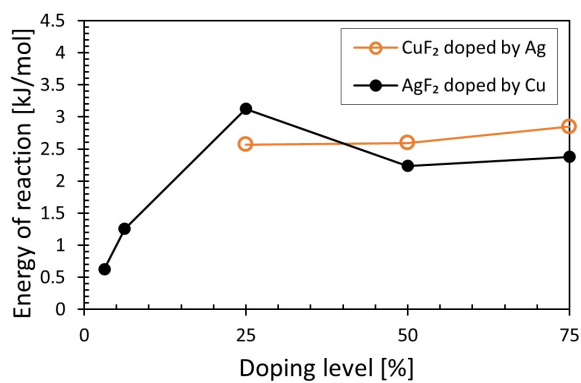

**Fig. SI5.1.** Energy of doping reaction, dE, for CuF$_2$ (Ag doping) and AgF$_2$ (Cu doping), defined as: dE = $E_{Ag_xCu_yF_2}$ - $(xE_{AgF_2} + yE_{CuF_2})$.

## SI6. DFT calculations for AgF$_2$/CuF$_2$ phases with disordered metal sites (Virtual Crystal Approximation)

**Table SI6.1.** Parameters of AgF$_2$-type cell upon the specific degree of Cu substitution in unordered approach (DFT, VCA). The "%diff" column is a percentage difference of specific parameter referring to pristine AgF$_2$ presented in column named as **0** (x value in orthorhombic Ag$_{1-x}$Cu$_x$F$_2$).

| Parameters | x - Cu concentration | | | | | | | | | |
|---|---|---|---|---|---|---|---|---|---|---|
| | 0 | 0.1 | % diff | 0.25 | % diff | 0.5 | % diff | 0.75 | % diff | 1 | % diff |
| a [Å] | 5.600 | 5.552 | -0.852 | 5.479 | -2.161 | 5.368 | -4.141 | 5.231 | -6.591 | 5.086 | -9.171 |
| b [Å] | 5.698 | 5.671 | -0.468 | 5.674 | -0.432 | 5.650 | -0.850 | 5.518 | -3.153 | 5.443 | -4.469 |
| c [Å] | 5.168 | 5.118 | -0.957 | 5.038 | -2.505 | 4.937 | -4.474 | 4.851 | -6.140 | 4.729 | -8.494 |
| V [Å$^3$] | 41.226 | 40.294 | -2.260 | 39.155 | -5.024 | 37.430 | -9.208 | 35.005 | -15.091 | 32.733 | -20.600 |

**Table SI6.2.** Parameters of CuF$_2$-type cell upon the specific degree of Ag substitution in unordered approach (DFT, VCA). The "%diff" column is a percentage difference of specific parameter referring to pristine CuF$_2$ presented in column named as **1** (x value in monoclinic Cu$_x$Ag$_{1-x}$F$_2$).

| Parameters | x - Cu concentration | | | | | | | | | |
|---|---|---|---|---|---|---|---|---|---|---|
| | 1 | 0.9 | % diff | 0.75 | % diff | 0.5 | % diff | 0.9 | % diff | 0 | % diff |
| a [Å] | 3.182 | 3.212 | +0.938 | 3.317 | +4.241 | 3.469 | +9.026 | 3.552 | +11.623 | 3.613 | +13.556 |
| b [Å] | 4.605 | 4.662 | +1.249 | 4.730 | +2.728 | 4.821 | +4.705 | 4.919 | +6.826 | 5.020 | +9.011 |
| c [Å] | 5.390 | 5.460 | +1.300 | 5.496 | +1.977 | 5.589 | +3.699 | 5.707 | +5.887 | 5.863 | +8.776 |
| V [Å$^3$] | 33.746 | 34.807 | +3.145 | 36.796 | +9.038 | 39.736 | +17.751 | 41.942 | +24.286 | 44.323 | +31.343 |

## SI7. Magnetic measurements

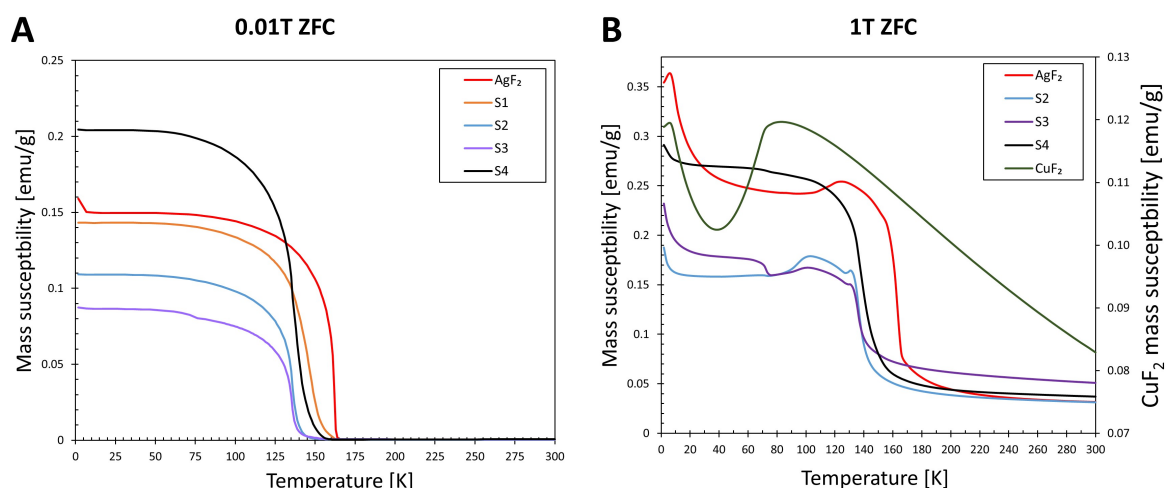

**Fig SI7.1.** (**A**) and (**B**) Mass susceptibility vs temperature. (**A**) **S1**-**S4** and $AgF_2$ at 0.01T; (**B**) for **S2**-**S4**, $AgF_2$ and $CuF_2$ at 1T field. Data for ZFC regime presented. In (**B**) measurement for **S1** was omitted due to no sign of $CuF_2$-type transition.

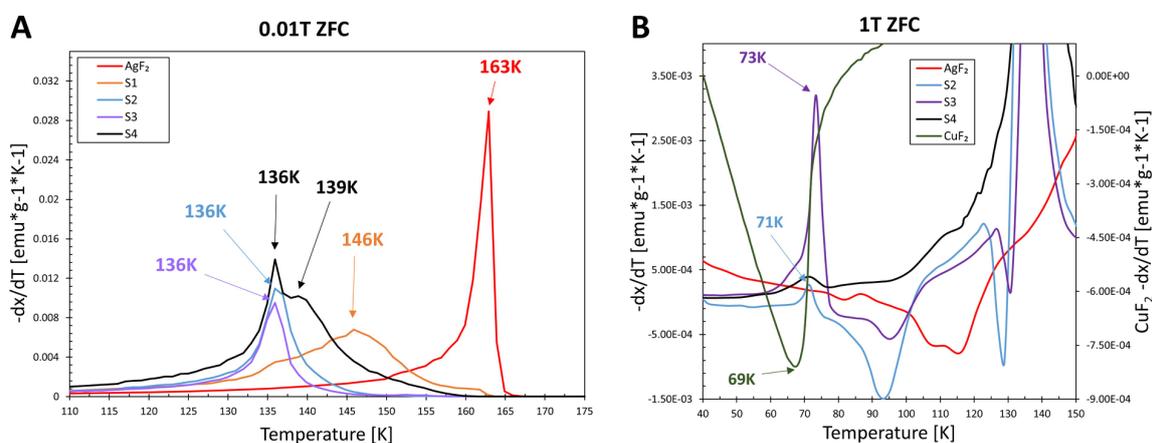

**Fig SI7.2.** (**A**) and (**B**) $-d\chi/dT$ vs temperature. (**A**) **S1**-**S4** and $AgF_2$ at 0.01 T at range 110-175 K; (**B**) for **S2**-**S4**, $AgF_2$ and $CuF_2$ at 1 T field, 40-150 K range. Data for ZFC regime are presented. In (**B**) measurement for **S1** was not shown as $CuF_2$-type transition could not be detected.

## SI8. Electronic impedance spectroscopy: ionic conductivity of sample S1

Impedance measurements were performed in the $10^{-2}$–$10^6$ Hz frequency range using a Solartron 1260 Frequency Response Analyzer with 1296 Dielectric Interface. Due to the high reactivity of samples, all impedance measurements, sample preparation, and loading were conducted inside a glovebox under an inert atmosphere. Measurements were carried out using our innovative setup IMPED CELL [3] (Fig.SI8.1). Stainless steel cylinders covered with nickel and passivated by nickel fluoride were used as electrodes. Such electrodes have sufficient hardness to compress the samples and chemical resistivity towards divalent silver and fluorine. The samples were compressed to 200 MPa directly between the electrodes with no binders or additional electric contacts. Thus, the samples were investigated in pure form. IMPED CELL monitors temperature, pressure, and sample thickness throughout the measurement, which allows reliable correction of the obtained results. The impedance was corrected to the resistivity of an empty cell (20 ohms). For each sample, several spectra were acquired, receiving consistent results. Nyquist plots of registered spectra have shapes of a deformed semicircle (**Fig. SI8.2.**). Due to the low conductivities of synthesized samples and commercial $AgF_2$ and $CuF_2$ the low-frequency part of the spectra was not visible. Only for AgF a spur at lower frequencies was registered, which is typical for materials exhibiting ionic conductivity. Fitting was carried out in ZPlot software from Scribner Associates, Inc. For fitting the spectra various forms of the dielectric function were tested: Cole-Cole, Davidson–Cole, Jonscher, and Havrilak Negami. The best fit quality was obtained for Havriliak–Negami model for ionic conductivity. The equivalent circuit used for fitting the spectra is presented in Fig. SI8.1. The equivalent circuit contains two branches, one related to high-frequency dielectric relaxation represented by a capacitor. The second is a series connection of Havriliak–Negami and if needed, the Warburg element.

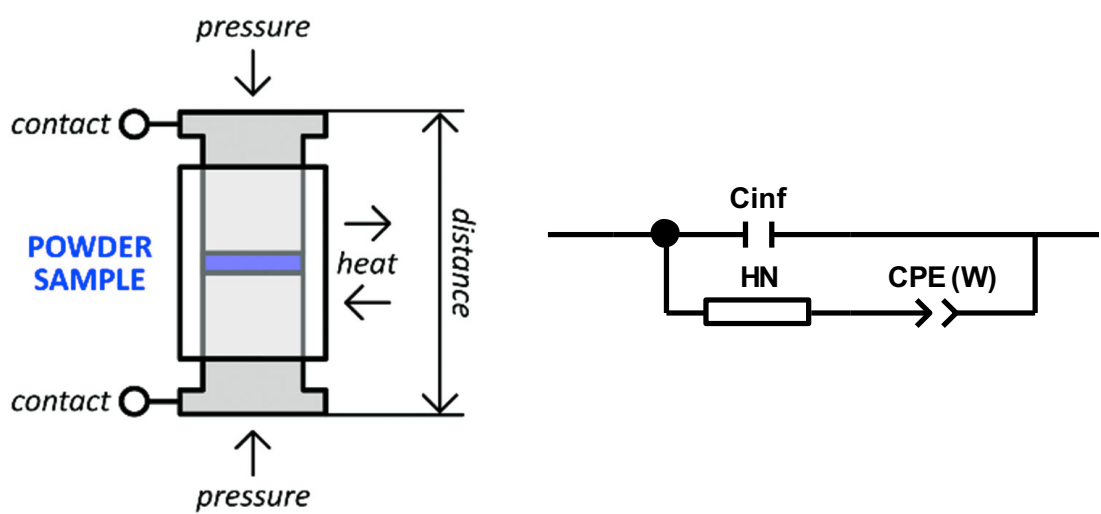

**Fig. SI8.1.** Scheme of IMPED CELL showing air-tight measurement module with indicated direction of compression and the equivalent circuit used for fitting EIS spectra.

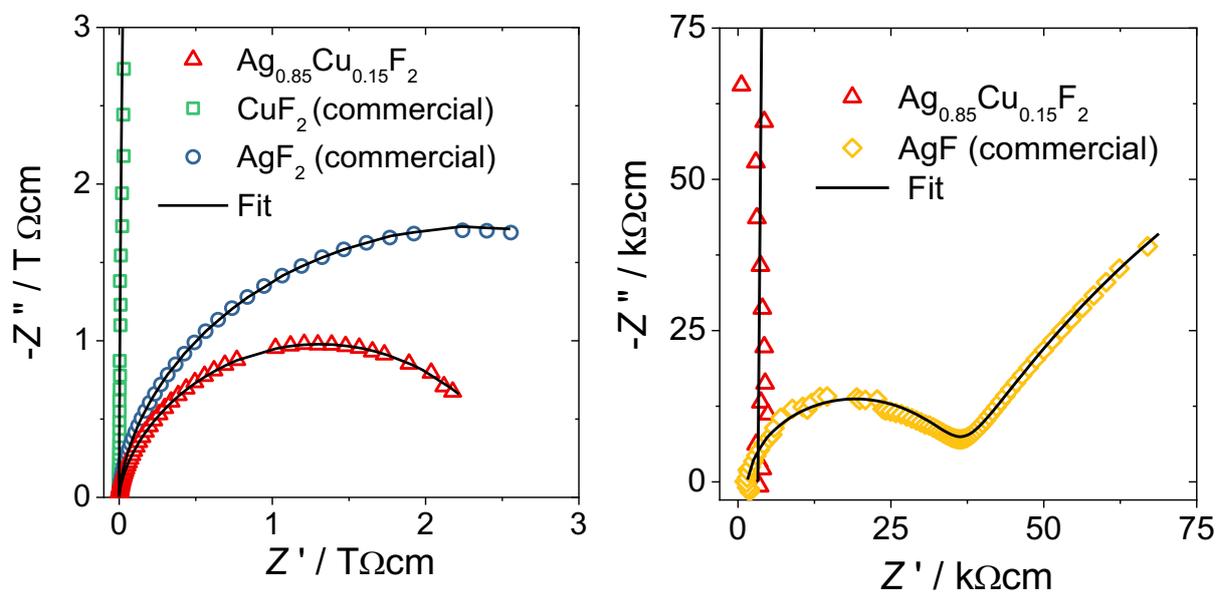

**Fig. SI8.2.** Nyquist plots of impedance spectra recorded in the temperature 20°C under 200 MPa normalized to the sample's effective thickens and electrode area.

**Table SI8.1.** Ionic conductivity of AgF, $AgF_2$, $CuF_2$ and $Ag_{0.85}Cu_{0.15}F_2$ (**S1**).

| # | Sample | ϭ [S/cm] |
|---|---|---|
| 1 | AgF commercial | $2.60 \cdot 10^{-5} \pm 1.6 \cdot 10^{-7}$ |
| 2 | $AgF_2$ commercial | $1.85 \cdot 10^{-13} \pm 1.2 \cdot 10^{-15}$ |
| 3 | $CuF_2$ commercial | $2.64 \cdot 10^{-15} \pm 1.6 \cdot 10^{-16}$ |
| 4 | $Ag_{0.85}Cu_{0.15}F_2$ (**S1**) | $3.71 \cdot 10^{-13} \pm 2.6 \cdot 10^{-15}$ |

# SI9. Theoretical DFT calculations of magnetic superexchange

Following the Ref[4-6] and assuming that Ising Hamiltonian can be written as a sum of the contribution from individual pairs of paramagnetic centers, we constructed system of equations of energies for each magnetic structure. In that convention positive sign of J refers to ferromagnetic ordering between metal centers, while negative value, to antiferromagnetic ordering. For mixed metal structures (50 % of doping), in order to mimic quasidisordered arrangement of Cu and Ag atoms in structures (like in solid solutions), we used different structures than those presented above in the section **SI.5** (see **Fig. SI9.2.** and **Fig. SI9.3**).

*$AgF_2$ – type structure*. To estimate J1, J2 and J3 values and to determinate the effect of Cu doping in $AgF_2$ we calculated energies of four magnetic spin states (see **Fig. SI9.1.**) for four structures: $AgF_2$, $CuF_2$ and two $Ag_{0.50}Cu_{0.50}F_2$ (see **Fig. SI9.2.**). Additionally, to determinate the influence of $U_{Ag}$ (Hubbard parameter) in DFT+U method on J constants, we used two different values: 5 eV and 8 eV in our calculations. Results are presented in **Table SI.9.1**.

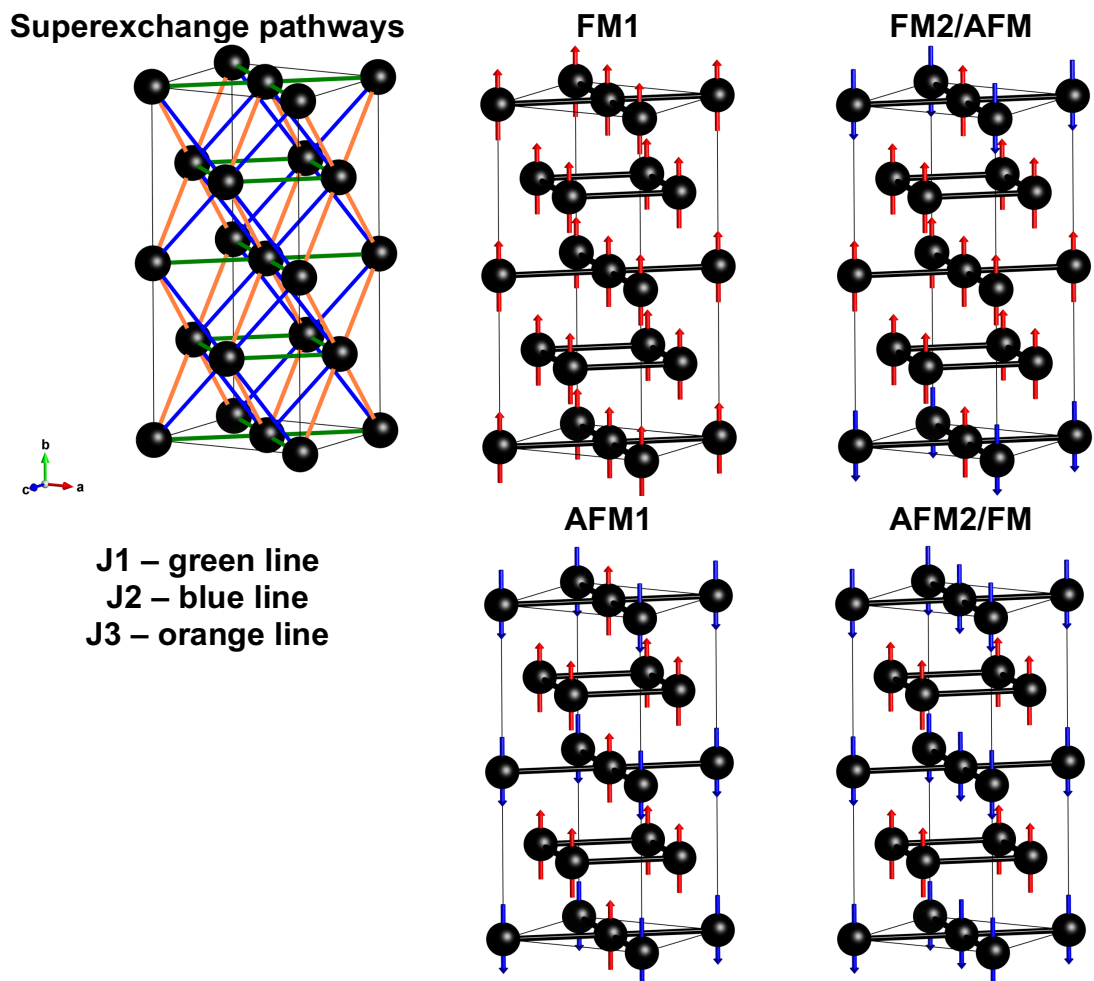

**Fig. SI9.1.** Superexchange pathways and four possible spin states for 1x2x1 $AgF_2$ supercell, following Ref. [7]. Fluorine atoms were omitted for clarity. Blue and red arrows represent opposite spin directions.

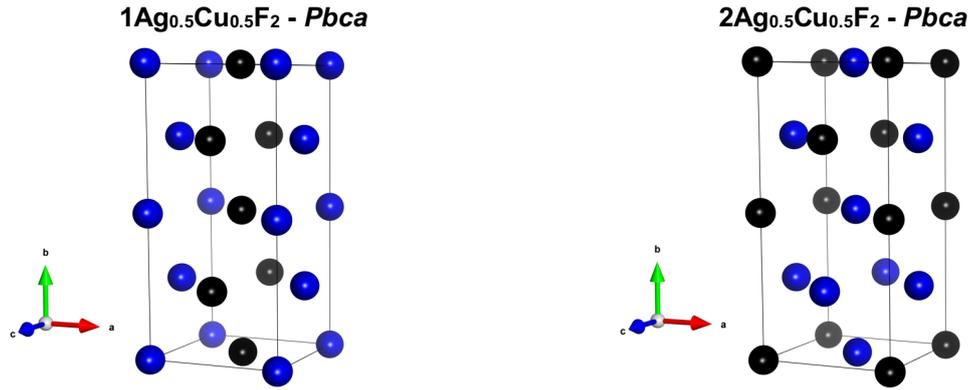

**Fig. SI9.2.** Two structures of $Ag_{0.5}Cu_{0.5}F_2$ adopted for magnetic calculations. Color code: Ag – black balls, Cu – blue balls.

**Table SI.9.1.** Superexchange constants (SE) values for $AgF_2$, two types of $Ag_{0.50}Cu_{0.50}F_2$ (see **Fig. SI9.1.** and **SI9.2.**) and $CuF_2$ (100% doping of $AgF_2$, *Pbca* type).

| SE constant | $U_{Ag}$ = 8 eV, $U_{Cu}$ = 10 eV | | | | $U_{Ag}$ = 5 eV, $U_{Cu}$ = 10 eV | | | |
|---|---|---|---|---|---|---|---|---|
| | *AgF₂* | *1Ag₀.₅Cu₀.₅F₂* | *2Ag₀.₅Cu₀.₅F₂* | *CuF₂* | *AgF₂* | *1Ag₀.₅Cu₀.₅F₂* | *2Ag₀.₅Cu₀.₅F₂* | *CuF₂* |
| **J1** | -34.22 | -16.21 | -15.81 | -6.98 | -51.67 | -19.10 | -18.29 | -6.98 |
| **J2** | 0.53 | 0.18 | 0.83 | 0.14 | 0.76 | 0.20 | 1.17 | 0.14 |
| **J3** | 1.50 | 0.86 | 0.20 | 0.33 | 3.31 | 1.34 | 0.40 | 0.33 |

All magnetic structures calculated in *Pbca* unit cells ($AgF_2$ – type). The J values in ordered $AgCuF_4$ with alternated metal-fluorine layers are presented in Ref. [8] (most stable polytype).

*$CuF_2$ – type structure.* To estimate J1, J2 and J3 values and to determinate the effect of Ag doping in $CuF_2$ we calculated energies of four magnetic spin states (see **Fig. SI9.1.**) for three structures: $CuF_2$, $Cu_{0.50}Ag_{0.50}F_2$, $AgF_2$ in *P21/c* unit cell (see **Fig. SI9.2.**). Additionally, to determinate the influence of $U_{Ag}$ (Hubbard parameter) in DFT+U method on J constants, we used two different values: 5 eV and 8 eV in our calculations. Results are presented in **Table SI.9.2.**

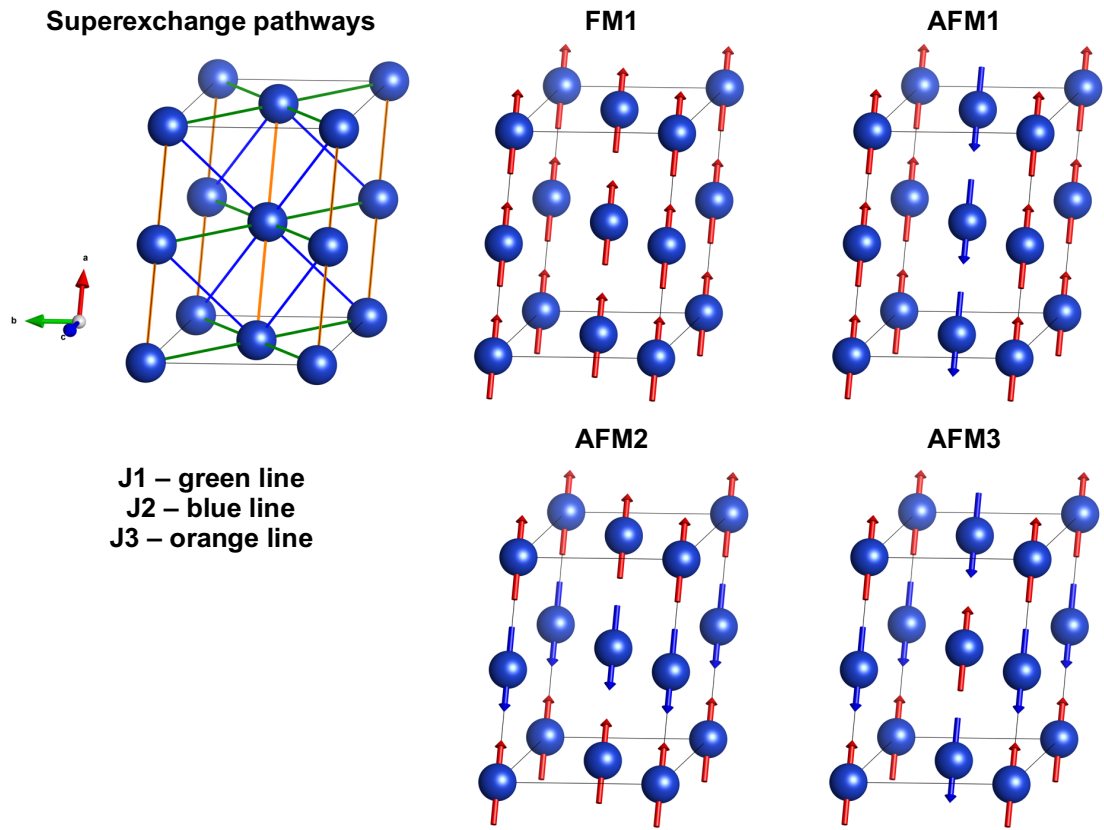

**Fig. SI9.3.** Superexchange pathways and four possible spin states for 2x1x1 $CuF_2$ supercell, following Ref. [9]. Fluorine atoms were omitted for clarity. Blue and red arrows represent opposite spin directions.

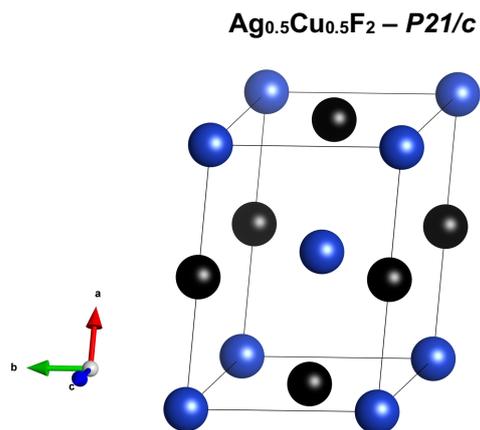

**Fig. SI9.4.** Structure of $Cu_{0.5}Ag_{0.5}F_2$ adopted for magnetic calculations. Color code: Ag – black balls, Cu – blue balls.

**Table SI.9.2.** Superexchange constants (SE) values for $CuF_2$, $Cu_{0.50}Ag_{0.50}F_2$ (see **Fig. SI9.3.** and **SI9.4.**) and $AgF_2$ (100% doping of $CuF_2$, *P21/c* type).

| SE constant | $U_{Ag}$ = 8 eV, $U_{Cu}$ = 10 eV | | | $U_{Ag}$ = 5 eV, $U_{Cu}$ = 10 eV | | |
|---|---|---|---|---|---|---|
| | *$CuF_2$* | *$Cu_{0.50}Ag_{0.50}F_2$* | *$AgF_2$* | *$CuF_2$* | *$Cu_{0.50}Ag_{0.50}F_2$* | *$AgF_2$* |
| J1 | -8.65 | -19.44 | -39.14 | -8.65 | -22.54 | -58.44 |
| J2 | 0.43 | 0.52 | 1.15 | 0.43 | 0.80 | 2.25 |
| J3 | -0.97 | -2.24 | -4.65 | -0.97 | -2.76 | -7.09 |

All magnetic structures calculated in *P21/c* unit cells ($CuF_2$ – type). Note, that the energy preferred structure is the one with alternation of [$AgF_2$] and [$CuF_2$] layers perpendicular to [100] plane (see **Table SI5.2.**), by -16.52 meV per FU.

## SI10. Milling of AgF$_2$ in nickel reactor

Mechanochemical reactions (high-energy milling) were carried out in milling bowls made of pure nickel metal. Millings were performed in 30-second periods with a 10 repetition in a series, using an L MW-S vibrational mill from Testchem (1400 rpm, ca. 23.3 Hz). The milling bowl was cooled between the periods, using liquid nitrogen (LN2) to avoid thermal decomposition and overheating of the products. Using this approach in each attempt about 200 mg of the commercial AgF$_2$ (>98%, Sigma) was put in the bowl in a dry argon gas atmosphere, which was then closed and moved to a vibrational mill. The samples of obtained products were sealed under the inert gas atmosphere inside 0.5 mm diameter quartz capillaries. XRD measurements were conducted using PANalytical X'Pert Pro diffractometer with a linear PIXcel Medipix2 detector (a parallel beam of CoKα$_1$ and CoKα$_2$, radiation intensity ratio of 2:1) at room temperature (**Fig. SI.10.1.**).

Measured XRD reflections are wide and have low intensity due to the small size of crystallites after the milling. Additionally, high noise and the presence of unknown crystal phases make Rietveld analysis difficult (**Fig. SI.10.1.**). Corrosion of the walls of nickel milling disk was observed (**Fig. SI10.2.**). However, the presence of known Ni-related phases in the powder sample is uncertain. XRD patterns either exclude the presence of polycrystalline AgNiF$_3$ or NiF$_2$, or indicate their content to be below the detection limit. Nevertheless, except AgF$_2$ (the starting material), AgF is present as a reduction product of silver difluoride. Nevertheless, except the AgF$_2$ (starting material), AgF presence can be noticed in samples **M1** and **M3**, while AgHF$_2$ only in **M3**, the two being products of silver difluoride reduction. Yet, neither of these was identified in sample **M2**. The remaining reflections are assigned to other AgF$_2$ decomposition products – since these reflections were identified also in other samples of AgF$_2$ where no nickel vessel was used – and are currently under examination.

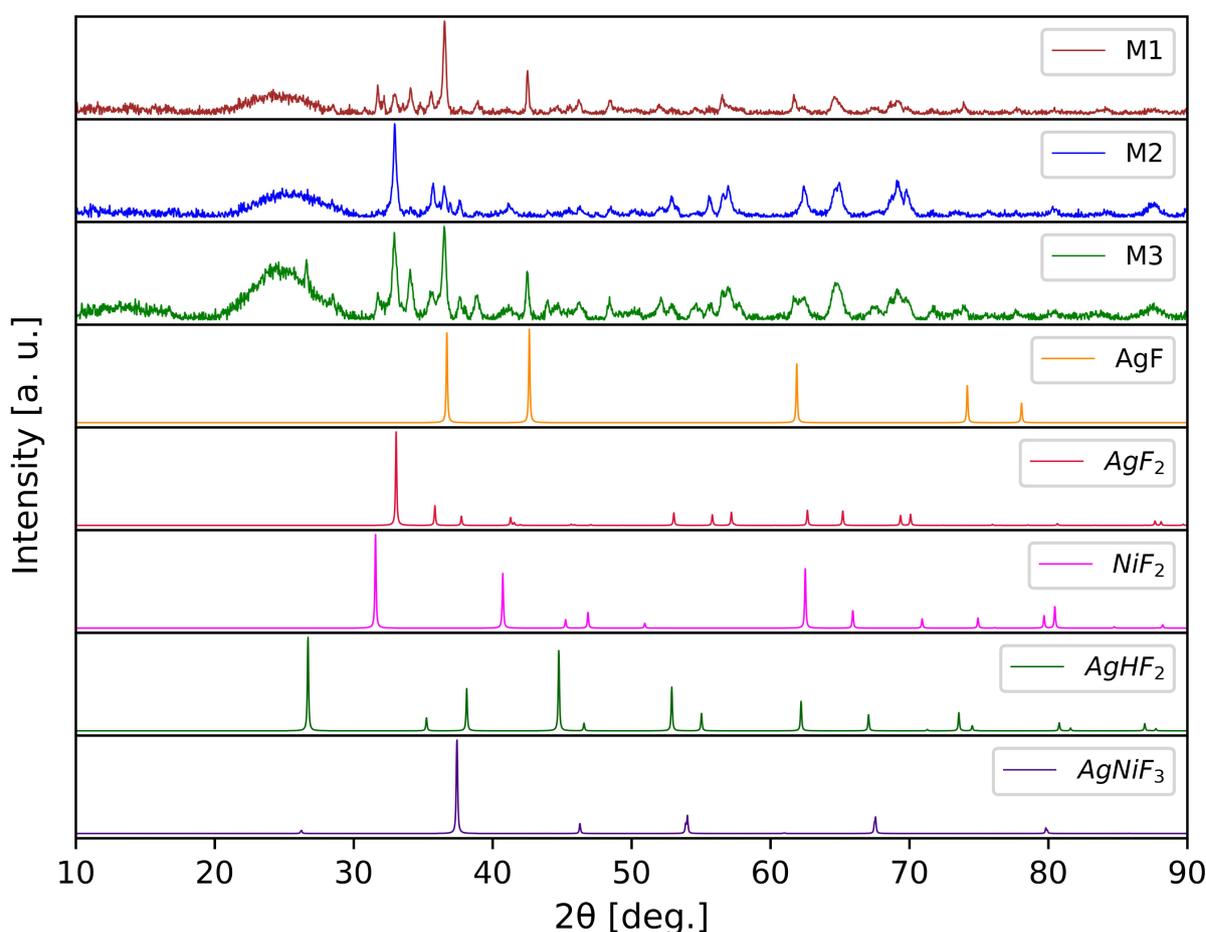

**Fig. SI10.1.** Powder X-ray diffraction (Co Kα radiation) patterns of three milling attempts (**M1**, **M2**, **M3**) with model patterns of AgF$_2$, AgF, NiF$_2$, AgHF$_2$ and AgNiF$_3$ as reference. Wide signal at 25° of 2θ comes from quartz capillary used for the measurements (remaining after background subtraction).

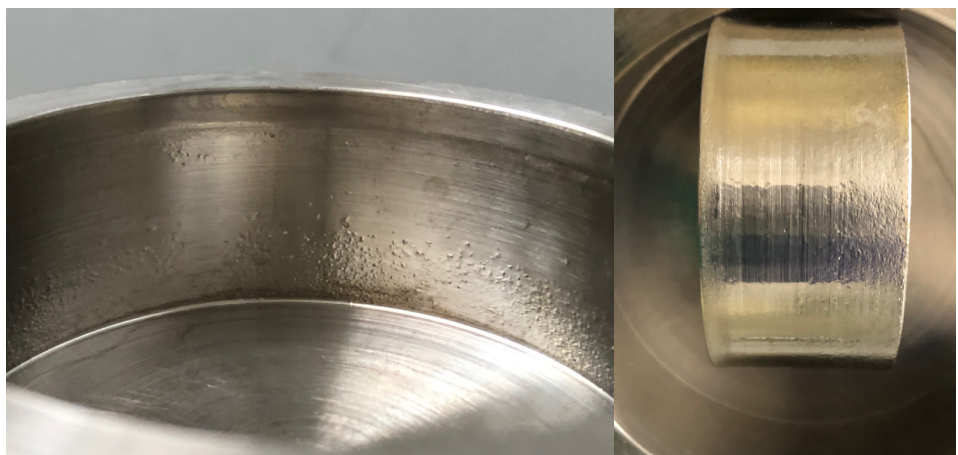

**Fig. SI10.2.** Corrosion of the nickel bowl after AgF$_2$ milling procedure.

## SI10. Bibliography

[1] H. D. B. Jenkins, L. Glasser, *Inorg. Chem.* **2003**, *42*, 8702.
[2] M. Domański, W. Grochala, *RSC Adv.* **2021**, *11*, 25801.
[3] K. Fijałkowski, P. Połczyński, R. Jurczakowski, European patent, EP 2 788 745.
[4] F. Illas, I. P. R. Moreira, C. de Graaf, V. Barone, *Theor. Chem. Acc.* **2000**, 104, 265.
[5] D. Dai, M.-H. Whangbo, *J. Chem. Phys.* **2001**, 114, 2887.
[6] D. Dai, M.-H. Whangbo, *J. Chem. Phys.* **2003**, 118, 29.
[7] D. Kurzydłowski, PhD dissertation, **2012**, Warsaw.
[8] M. Domański, W. Grochala, *Z. Naturforsch. B* **2021**, *76*, 751.
[9] P. Reinhardt, I. de P. R. Moreira, C. de Graaf, R. Dovesi, F. Illas, *Chem. Phys. Lett.* **2000**, *319*, 625.